\documentclass[aps,prd,twocolumn,showpacs,superscriptaddress,groupedaddress]{revtex4-1}  
\usepackage[colorlinks,linkcolor=blue,anchorcolor=blue,urlcolor=blue,citecolor=blue,bookmarks,bookmarksnumbered]{hyperref}
\usepackage{graphicx}  
\usepackage{dcolumn}   
\usepackage{bm}        
\usepackage{amssymb}   
\usepackage{amsmath}
\usepackage{multirow}
\usepackage{longtable}
\usepackage{float}
\usepackage{bbding}
\newcommand{\TIMES}{\times}
\usepackage{enumitem}
\usepackage{tikz}
\usepackage{etoolbox}
\newcommand{\circled}[2][]{\tikz[baseline=(char.base)]
    {\node[shape = circle, draw, inner sep = .0pt]
    (char) {\phantom{\ifblank{#1}{#2}{#1}}};%
    \node at (char.center) {\makebox[0pt][c]{#2}};}}
\robustify{\circled}
\newcommand{\cirn}{\circled[10]}

\begin{document}

\newcommand*{\PKU}{School of Physics and State Key Laboratory of Nuclear Physics and
Technology, Peking University, Beijing 100871,
China}\affiliation{\PKU}
\newcommand*{\CICQM}{Collaborative Innovation
Center of Quantum Matter, Beijing, China}\affiliation{\CICQM}
\newcommand*{\CHEP}{Center for High Energy
Physics, Peking University, Beijing 100871,
China}\affiliation{\CHEP}

\title{Examination of pairs in neutrino mixing matrix}

\author{Dianjing Liu}\affiliation{\PKU}
\author{Bo-Qiang Ma}\email{mabq@pku.edu.cn}\affiliation{\PKU}\affiliation{\CICQM}\affiliation{\CHEP}

\begin{abstract}
  We examine the pairs of neutrino mixing matrix and suggest pairs that can be used in the construction of new mixing patterns, with ``pair" denoting the equality
  of the modulus of a pair of matrix elements. The results show that the trimaximal mixing in $\nu_2$ and the $\mu$-$\tau$ interchange symmetry are good choices under current experimental results. The two cases of bipair mixing pattern depend on the mass hierarchy of neutrinos. We also derive constraints on the {\it CP} phase by the pairs. The results are compatible with the maximal {\it CP} violation in most cases that are both self-consistent and consistent with experimental results.
\end{abstract}

\pacs{}
\maketitle

\section{Introduction}\label{sec:Introduction}

It has been firmly established that neutrinos can transit from one flavor to another from various oscillation experiments. In the
framework of three-generation neutrinos, the neutrino mass eigenstates are connected to flavor eigenstates by a unitary matrix, i.e., the
Pontecorvo-Maki-Nakagawa-Sakata~(PMNS) matrix~\cite{PMNS}. In the standard parametrization, i.e., the Chau-Keung~(CK) scheme~\cite{CK}, the PMNS matrix is
expressed as
\begin{widetext}
\begin{equation}
U_{\textrm{PMNS}}=
\left(
\begin{array}{ccc}
c_{12}c_{13}&s_{12}c_{13}&s_{13}e^{-i\delta}\\
-s_{12}c_{23}-c_{12}s_{23}s_{13}e^{i\delta}&c_{12}c_{23}-s_{12}s_{23}s_{13}e^{i\delta}&s_{23}c_{13}\\
s_{12}s_{23}-c_{12}c_{23}s_{13}e^{i\delta}&-c_{12}s_{23}-s_{12}c_{23}s_{13}e^{i\delta}&c_{23}c_{13}
\end{array}
\right)
\left(
\begin{array}{ccc}
1&&\\ &e^{i\alpha}&\\ &&e^{i\beta}
\end{array}
\right) ,
\label{PMNS-matrix}
\end{equation}
\end{widetext}
where $s_{ij}$ denotes $\sin\theta_{ij}$ and $c_{ij}$ denotes $\cos\theta_{ij}$ ($i,j=1,2,3$). The phase matrix $\mathrm{Diag}\{1,e^{i\alpha},e^{i\beta}\}$ denotes the contribution from the Majorana type neutrinos and the two phases $\alpha$ and $\beta$ do not manifest themselves in oscillations. Thus, there remain three mixing angles ($\theta_{12},\theta_{13},\theta_{23}$) and a {\it CP} phase $\delta$ for the description of neutrino oscillations. While experimental data on three mixing angles have been coming out continuously, there is no direct experimental measurement on $\delta$. Nevertheless, some indirect analyses, including analysis of experiments on reactor and accelerator neutrinos~\cite{analysis} and global fit results~\cite{global}, suggest that the {\it CP} phase is close to $-90^\circ$ (assuming $\delta\in[-180^\circ,180^\circ]$). This is in accord with the maximal {\it CP} violation, i.e., $\delta=\pm90^\circ$.

On the other hand, the search of mixing patterns of the PMNS matrix is a way to understand properties of neutrinos. In the search of mixing patterns, the concept of ``pair'' is often used, with a ``pair'' referring to the equality of the modulus of a pair of matrix elements. For example, the long discussed trimaximal mixing in $\nu_2$~\cite{tm} can be expressed as three pairs
\begin{equation}
|U_{12}|=|U_{22}|,
\end{equation}
\begin{equation}
|U_{12}|=|U_{32}|,
\end{equation}
\begin{equation}
|U_{22}|=|U_{32}|,
\end{equation}

where $U_{ij}$ denotes the corresponding element (with row $i$ and column $j$) of the PMNS matrix, i.e., Eq.~(\ref{PMNS-matrix}).
Similarly the $\mu$-$\tau$ interchange symmetry~\cite{mutau1,mutau2} can be expressed as
\begin{eqnarray}
|U_{21}|=|U_{31}|,\\ \nonumber\\
|U_{22}|=|U_{32}|,\\ \nonumber\\
|U_{23}|=|U_{33}|.
\end{eqnarray}
The so-called bipair mixing~\cite{bp} assigns
\begin{eqnarray}
|U_{12}|=|U_{32}|,\\ \nonumber\\
|U_{22}|=|U_{23}|,
\end{eqnarray}
as case (1), and
\begin{eqnarray}
|U_{12}|=|U_{22}|,\\ \nonumber\\
|U_{32}|=|U_{33}|,
\end{eqnarray}
as case (2).

These phenomenological relations are included in many mixing patterns. An example is the extensively studied tri-bimaximal mixing pattern~(TBM)~\cite{tbm}
\begin{equation}
U_{\textrm{TBM}}=\left(
\begin{array}{ccc}
\sqrt{\frac23}&\frac1{\sqrt3}&0\\
-\frac1{\sqrt6}&\frac1{\sqrt3}&\frac1{\sqrt2}\\
\frac1{\sqrt6}&-\frac1{\sqrt3}&\frac1{\sqrt2}\\
\end{array}
\right)
,
\end{equation}
which includes the trimaximal mixing and the $\mu$-$\tau$ symmetry, as well as the assumption of $|U_{13}|=0$.
Another one is the bimaximal mixing pattern~(BM)~\cite{bm}
\begin{equation}
U_{\textrm{BM}}=\left(
\begin{array}{ccc}
\frac1{\sqrt2}&\frac1{\sqrt2}&0\\
-\frac12&\frac12&\frac1{\sqrt2}\\
\frac12&-\frac12&\frac1{\sqrt2}
\end{array}
\right)
,
\end{equation}
which includes the $\mu$-$\tau$ symmetry, the assumption of $|U_{13}|=0$ and a pair $|U_{11}|=|U_{12}|$.
The bipair mixing pattern~\cite{bp}, originally based on bipair mixing and $|U_{13}|=0$, is described as
\begin{equation}
U_{\textrm{BP}}=\left(
\begin{array}{ccc}
c_{12}&s_{12}&0\\
-t_{12}^2&t_{12}&t_{12}\\
s_{12}t_{12}&-s_{12}&t_{12}/c_{12}
\end{array}
\right)
,
\end{equation}
for case (1), and
\begin{equation}
U_{\textrm{BP}}=\left(
\begin{array}{ccc}
c_{12}&s_{12}&0\\
-s_{12}t_{12}&s_{12}&t_{12}/c_{12}\\
t_{12}^2&-t_{12}&t_{12}
\end{array}
\right)
,
\end{equation}
for case (2), with $s_{12}^2=1-1/\sqrt{2}$.

Although the hypothesis $|U_{13}|=0$ contradicts the new data from the accelerator and reactor neutrino oscillation experiments~\cite{experiment}, other relations based on the pairs may still hold. For example, a new mixing pattern~\cite{newpattern} was proposed based on a nonzero $\theta_{13}$, the $\mu$-$\tau$ symmetry, the trimaximal mixing and the self-complementarity relation~\cite{selfcomplementarity} (i.e. $\theta_1+\theta_3=45^\circ$ in another parametrization scheme). The new pattern has the form
\begin{equation}
|U|=\left(
\begin{array}{ccc}
\frac{\sqrt2+1}3&\frac1{\sqrt3}&\frac{\sqrt2-1}3\\
\frac{\sqrt{3-\sqrt2}}3&\frac1{\sqrt3}&\frac{\sqrt{3+\sqrt2}}3\\
\frac{\sqrt{3-\sqrt2}}3&\frac1{\sqrt3}&\frac{\sqrt{3+\sqrt2}}3
\end{array}
\right),
\end{equation}
and it makes a prediction of maximal {\it CP} violation $\delta=\pm90^\circ$.

Therefore, the pairs would help in the search of new mixing patterns of the PMNS matrix. In order to know which pairs to choose in the construction of new mixing patterns, it is worthwhile to examine which of the pairs in the PMNS matrix are consistent with current experimental results, and whether they are consistent with each other.

What is more, the introduction of each pair produces a constraint on the four parameters of the PMNS matrix. Together with global fit results on three mixing angles, each pair would give a range of the {\it CP} phase (as is discussed in Sec.~\ref{sec:singlepair}, there are some exceptions in which the constraints do not include $\delta$). Examinations of these ranges would give information about the consistency among pairs and the consistency between the pairs and the global fit results.

In Sec.~\ref{sec:singlepair} we consider constraints by each single pair separately. By comparing the pair constraints with the natural limit of $\cos\delta$, we evaluate their consistency with global fit results.
In Sec.~\ref{sec:multipair} we combine ranges of the pairs to give joint constraints, and discuss their self-consistency and consistency with global fit results.
In Sec.~\ref{sec:mCPV} we pick out cases that are self-consistent and consistent with experimental results, and compare their ranges to the maximal {\it CP} violation.
Section \ref{sec:conclusion} is served for conclusions.

\section{single pair constraints}\label{sec:singlepair}

In our article all ranges of $\delta$ come from the ranges of $\cos\delta$. Thus we only discuss on the assumption that $\delta\in[0^\circ,180^\circ]$. When extending to $[-180^\circ,180^\circ]$, the results of $\delta\in[\delta_1,\delta_2]$ should also be extended to be $\delta\in[-\delta_2,-\delta_1]$ and $[\delta_1,\delta_2]$.

We consider a single pair, for example, $|U_{21}|=|U_{31}|$. It gives rise to a relation between $\cos\delta$ and mixing angles. In the case of $|U_{21}|=|U_{31}|$, it is \\
\begin{widetext}
\begin{equation}
\cos\delta(\theta_{12},\theta_{13},\theta_{23})=\frac{(\sin\theta_{12}\sin\theta_{23})^2+(\sin\theta_{13}\cos\theta_{12}\cos\theta_{23})^2-(\sin\theta_{12}\cos\theta_{23})^2-(\sin\theta_{13}\sin\theta_{23}\cos\theta_{12})^2}{4\sin\theta_{12}\sin\theta_{13}\sin\theta_{23}\cos\theta_{12}\cos\theta_{23}}.
\end{equation}
\end{widetext}

By inserting the global fit results of mixing angles~\cite{global}
\begin{equation}
\theta_{12}=(33.48^{+0.78}_{-0.75})^\circ,
\end{equation}
\begin{equation}
\theta_{13}=(8.50^{+0.20}_{-0.21})^\circ\oplus(8.51^{+0.20}_{-0.21})^\circ,
\end{equation}
\begin{equation}
\theta_{23}=(42.3^{+3.0}_{-1.6})^\circ\oplus(49.5^{+1.5}_{-2.2})^\circ,
\end{equation}
we obtain the central value and $1\sigma$ error of $\delta$, which are
\begin{equation}
\cos\delta=-0.2009^{+0.2247}_{-0.1201}
\end{equation}
for normal hierarchy (NH), and
\begin{equation}
\cos\delta=0.3362^{+0.1148}_{-0.1677}
\end{equation}
for inverted hierarchy (IH).

\begin{figure}[H]
  \includegraphics[width=.48\linewidth]{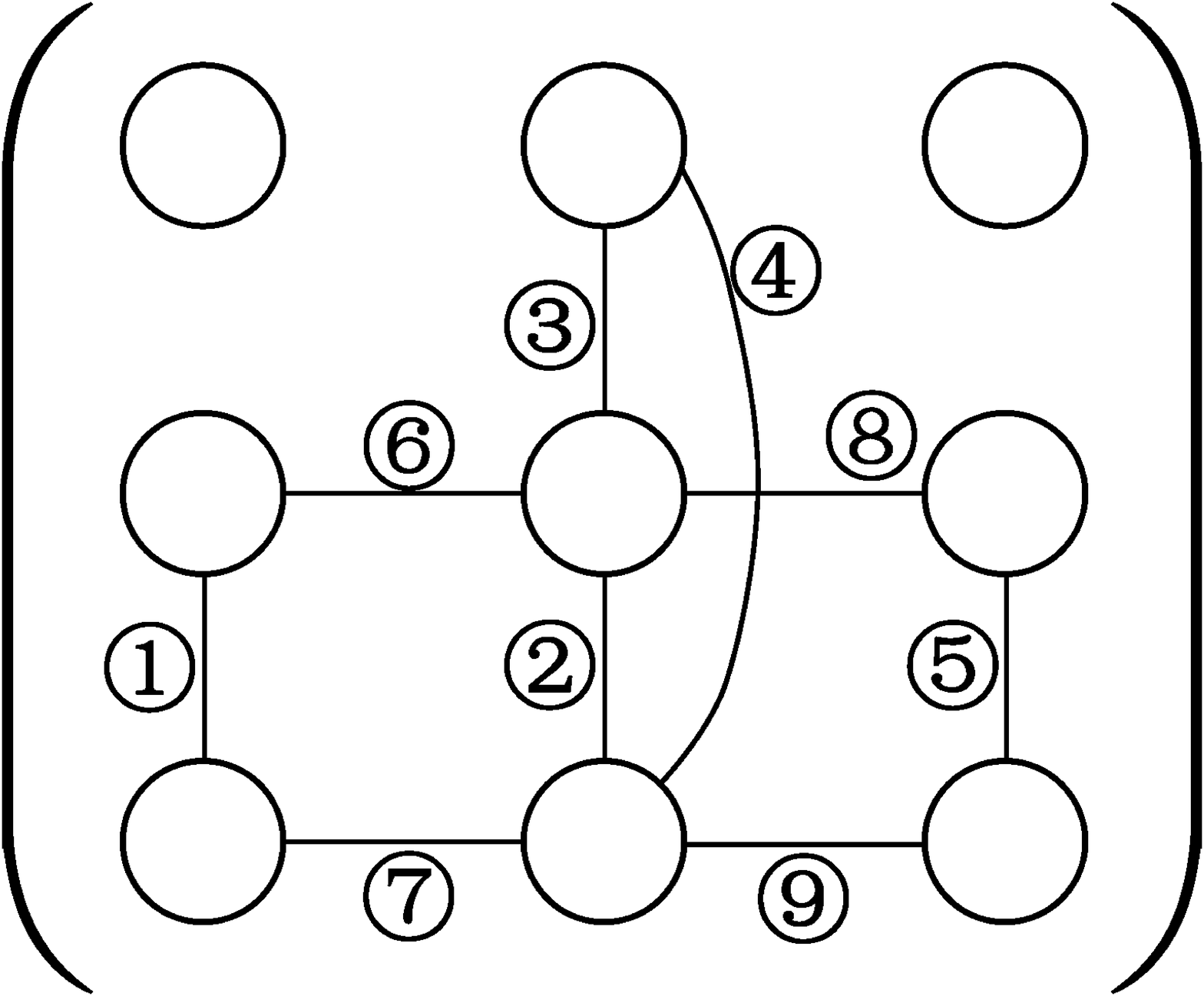}
  \includegraphics[width=.48\linewidth]{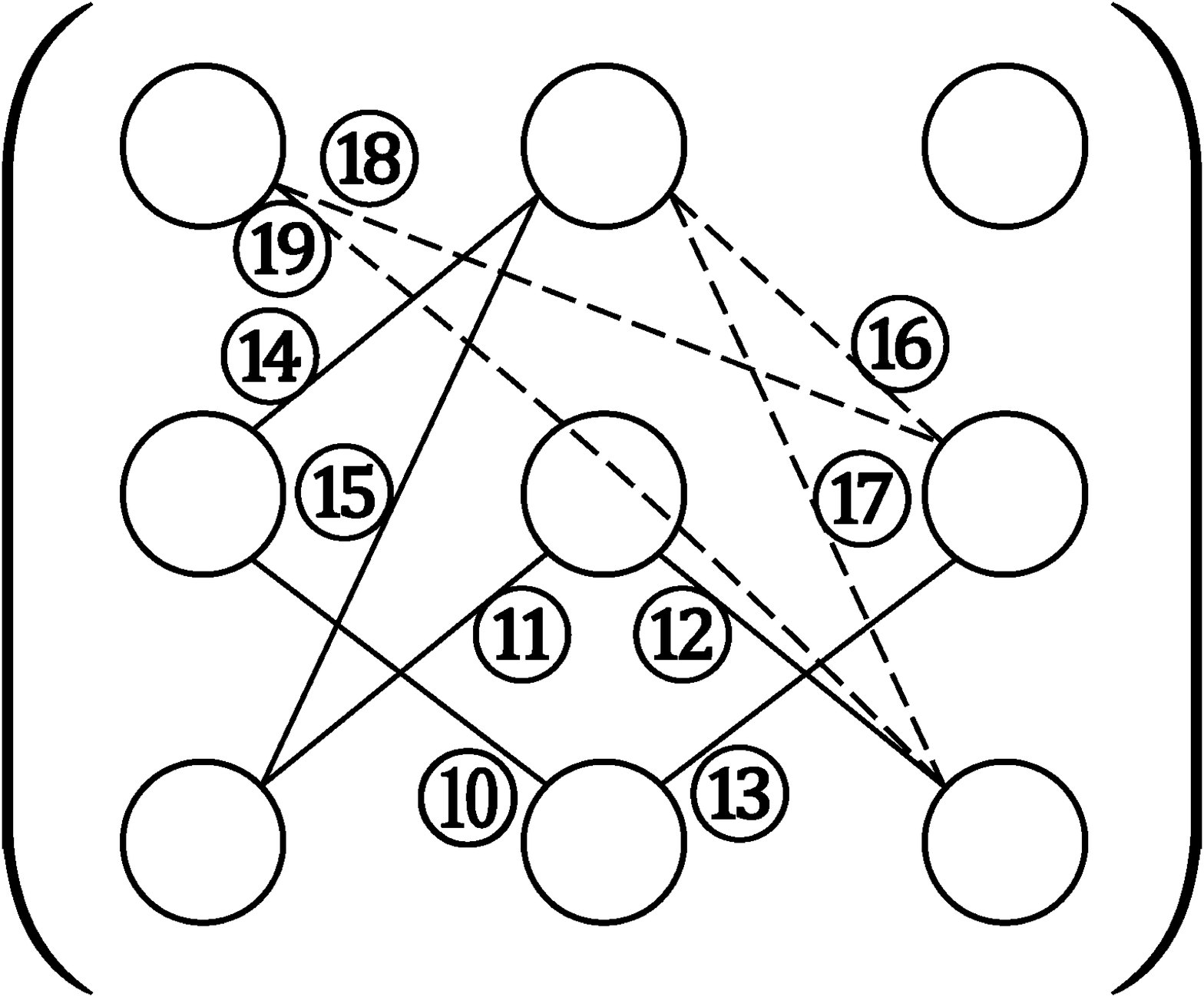}
  \caption{\label{fig:pairnum}
  The identification of possible pairs with numbers \cirn{\small{1}}-\cirn{\small{19}}. Pairs \cirn{\small{1}}-\cirn{\small{15}} have overlap in $3\sigma$ range. Pairs \cirn{\small{16}}-\cirn{\small{19}} have overlap in $4\sigma$ range instead of $3\sigma$. They are denoted by dash lines but also included in discussion.}
\end{figure}

From global fit of $3\sigma$ ranges on the magnitude of matrix elements~\cite{global}
\begin{widetext}
\begin{equation}
|U|=\left(
\begin{array}{ccc}
0.801\rightarrow0.845&0.514\rightarrow0.580&0.137\rightarrow0.158\\
0.225\rightarrow0.517&0.441\rightarrow0.699&0.614\rightarrow0.793\\
0.246\rightarrow0.529&0.464\rightarrow0.713&0.590\rightarrow0.776
\end{array}
\right), \label{eq5}
\end{equation}
\end{widetext}
we pick out all pairs whose corresponding matrix elements have overlap in $3\sigma$ range, and number them as in Fig.~\ref{fig:pairnum}. Pairs \cirn{\small{1}}-\cirn{\small{15}} have overlap in $3\sigma$ range. Pairs \cirn{\small{16}}-\cirn{\small{19}} have overlap in $4\sigma$ range instead of $3\sigma$. Although pairs \cirn{\small{16}}~\cirn{\small{19}} (or \cirn{\small{17}}~\cirn{\small{18}}) do not hold in $3\sigma$ range and therefore need not be discussed by principle, their constraints on mixing angles are the same as that of pair \cirn{\small{10}} (or \cirn{\small{11}}). Thus, they are also included in the discussion.

\begin{figure*}[t]
  \includegraphics[width=\linewidth]{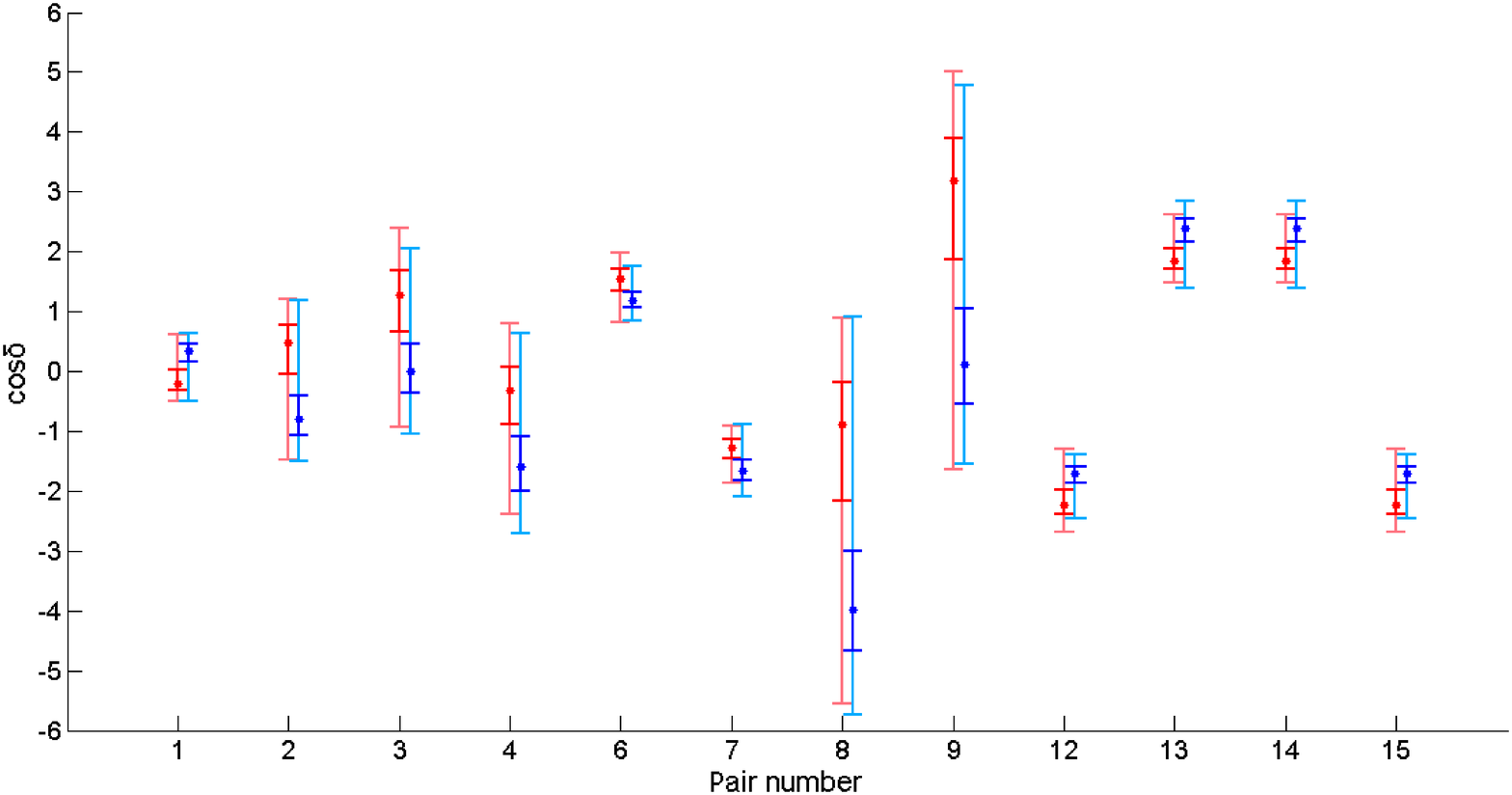}
  \caption{\label{fig:cosdelta}
  Constraints on $\cos\delta$ produced by single pair. Each error bar shows $1\sigma$ and $3\sigma$ ranges of $\cos\delta$. The left and right error bars of each pair denote ranges
  in NH and IH respectively.}
\end{figure*}

Similar to previous calculations, each pair produces a constraint on $\cos\delta$, except pairs \cirn{\small{5}}~\cirn{\small{10}}~\cirn{\small{11}}~\cirn{\small{16}}~\cirn{\small{17}}~\cirn{\small{18}}~\cirn{\small{19}}. Among them, pair \cirn{\small{5}} yields
\begin{equation}
\theta_{23}=45^\circ,
\end{equation}
\cirn{\small{10}}, \cirn{\small{16}}, \cirn{\small{19}} yield the same relation
\begin{equation}
\theta_{12}=\theta_{23},
\end{equation}
and \cirn{\small{11}},  \cirn{\small{17}}, \cirn{\small{18}} all yield
\begin{equation}
\theta_{12}+\theta_{23}=90^\circ.
\end{equation}
Other results are shown in Fig.~\ref{fig:cosdelta}.

In addition, pairs \cirn{\small{12}} and \cirn{\small{15}} produce the same relation between $\delta$ and mixing angles, as well as \cirn{\small{13}} and \cirn{\small{14}}. That is
\begin{eqnarray}
\cos\delta_{\circled[4]{\tiny{12}}}(\theta_{12},\theta_{13},\theta_{23})=\cos\delta_{\circled[4]{\tiny{15}}}(\theta_{12},\theta_{13},\theta_{23}),\\ \nonumber\\
\cos\delta_{\circled[4]{\tiny{13}}}(\theta_{12},\theta_{13},\theta_{23})=\cos\delta_{\circled[4]{\tiny{14}}}(\theta_{12},\theta_{13},\theta_{23}).
\end{eqnarray}
Therefore, \cirn{\small{12}} and \cirn{\small{15}} would produce the same constraint on $\delta$, as well as \cirn{\small{13}} and \cirn{\small{14}}.

While pair \cirn{\small{1}} gives a strong constraint on $\cos\delta$ with the natural limit $\cos\delta\in[-1,1]$ satisfied as well, it is necessary to point out that some constraints are completely not in agreement with the natural limit. Since the constraints are produced by pairs together with experimental results, it indicates that these pairs are not so consistent with current experimental results. Thus they might not be good choices when considering a new mixing pattern.

We divide the pairs into several classes according to the level of consistency between their constraints and the natural limit of $\cos\delta$. These classifications can be regarded as indications of their consistency with the experimental results. The results are shown in Table~\ref{tab:pairclass}.
\begin{table*}
  \caption{\label{tab:pairclass}
  The classification of pairs according to their ranks of consistency (marked with \FiveStarOpen) with the natural limit $\cos\delta\in[-1,1]$. While pairs \cirn{\small{5}}~\cirn{\small{10}}~\cirn{\small{11}}~\cirn{\small{16}}~\cirn{\small{17}}~\cirn{\small{18}}~\cirn{\small{19}} have no constraints on $\cos\delta$, they are divided according to their constraints on mixing angles and their ranks are denoted by $\bullet$.}
  \begin{ruledtabular}
  \begin{tabular}{cccc}
  \multirow{2}{*}{Rank of consistency}&\multirow{2}{*}{Constraint compared with natural limit $\cos\delta\in[-1,1]$}&\multicolumn{2}{c}{Pair}\\
  \cline{3-4}
  &&Normal hierarchy&Inverted hierarchy
  \\ \hline
  5\FiveStarOpen&$1\sigma,3\sigma$ within [-1,1]&\cirn1&\cirn1 \\[.1cm]
  4\FiveStarOpen&$1\sigma$ within [-1,1], $3\sigma$ partially beyond &\cirn2~\cirn4&\cirn3\\[.1cm]
  3\FiveStarOpen&Central value within [-1,1], $1\sigma$ partially beyond&\cirn8&\cirn2~\cirn9\\[.1cm]
  2\FiveStarOpen&Central value beyond [-1,1], $1\sigma$ partially within &\cirn3&$\cdot\cdot\cdot$\\[.1cm]
  1\FiveStarOpen&$1\sigma$ beyond [-1,1], $3\sigma$ partially within &\cirn6~\cirn7~\cirn9 &\cirn4~\cirn6~\cirn7~\cirn8\\[.1cm]
  0\FiveStarOpen&$3\sigma$ beyond [-1,1]&\circled{12}~\circled{13}~\circled{14}~\circled{15} &\circled{12}~\circled{13}~\circled{14}~\circled{15}\\[.1cm]
  $\bullet$&No constraint on $\cos\delta$, $\theta_{23}=45^\circ$ &\cirn5 &\cirn5 \\[.1cm]
  $\bullet$&No constraint on $\cos\delta$, $\theta_{12}=\theta_{23}$&\circled{10}~\circled{16}~\circled{19}&\circled{10}~\circled{16}~\circled{19}\\[.1cm]
  $\bullet$&No constraint on $\cos\delta$, $\theta_{12}+\theta_{23}=90^\circ$&\circled{11}~\circled{17}~\circled{18}&\circled{11}~\circled{17}~\circled{18}\\
  \end{tabular}
\end{ruledtabular}\end{table*}

\begin{figure*}[t]
  \includegraphics[width=\linewidth]{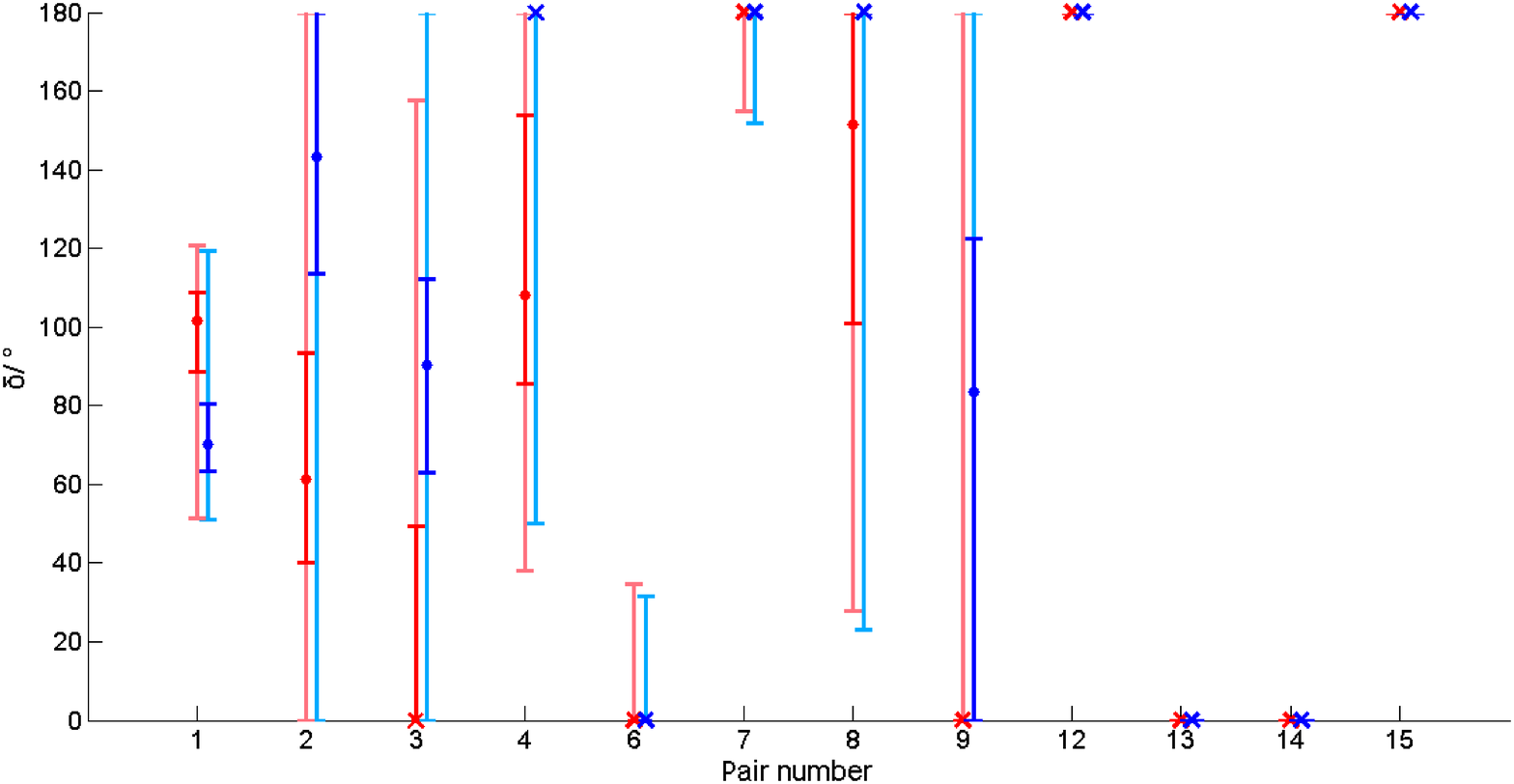}
  \caption{\label{fig:delta}Constraints on $\delta$ produced by single pair and the natural limit $\cos\delta\in[-1,1]$. Each error bar shows $1\sigma$ and $3\sigma$ ranges of $\delta$. The left and right error bars of each pair denote ranges in NH and IH respectively. The $\times$ means that the central value is out of physical range.}
\end{figure*}

Together with the natural limit, each of the pairs except \cirn{\small{5}}~\cirn{\small{10}}~\cirn{\small{11}}~\cirn{\small{16}}~\cirn{\small{17}}~\cirn{\small{18}}~\cirn{\small{19}} gives a constraint on $\delta$. The ranges of $\delta$ on the assumption that $\delta\in[0^\circ,180^\circ]$ are shown in Fig.~\ref{fig:delta}.

\section{combined pair constraints}\label{sec:multipair}
As is discussed in Sec.~\ref{sec:singlepair}, each pair produces a central value and an error of $\cos\delta$. If some of the pairs are supposed to hold simultaneously, their ranges should be regarded as measurements of the same Gaussian distribution of $\cos\delta$ and should be combined to give an estimation of the distribution.

When combining ranges from different pairs, we adopt standard weighted least-squares procedure advocated by the Particle Data Group~\cite{pdg}. The weighted average and error are
\begin{equation}
\overline{\cos\delta}\pm\sigma_{\overline{\cos\delta}}=\frac{\Sigma_i\omega_i\cos\delta_i}{\Sigma_i\omega_i}\pm(\Sigma_i\omega_i)^{-\frac12} , \label{eq1}
\end{equation}
where
\begin{equation}
\omega_i=\frac{1}{\sigma_{(\cos\delta_i)}^2} , \label{eq2}
\end{equation}
with $i$ referring to pairs which are combined.

The scale factor is defined as
\begin{equation}
S=\sqrt{\frac{\chi^2}{N-1}}, \label{eq3}
\end{equation}
whose expectation value is 1 since the expectation value of $\chi^2$ is $N-1$.

For cases with $S>1$, we also calculate scaled output errors, which are
\begin{equation}
\sigma_{\textrm{scaled}}=S\sigma_{\textrm{unscaled}}.\label{eq4}
\end{equation}
The reason is as follows: the relatively large value of $\chi^2$ is likely to be due to underestimation of errors in some of the measurements. Not knowing which of the input errors are underestimated, we assume they are all underestimated by the same factor $S$. If we scale up all the input errors by $S$, the $\chi^2$ becomes $N-1$, and the output error scales up by the same factor.

What is more, the p-value of the combination is calculated (with unscaled input errors), and the corresponding confidence level serves to indicate exclusion level about self-consistency of the combination.

\subsection{$\mu$-$\tau$ symmetry and trimaximal mixing}

First we combine pairs from the trimaximal mixing (i.e., pairs \cirn{\small{2}}~\cirn{\small{3}}~\cirn{\small{4}}) to explore a joint constraint on $\cos\delta$ by this phenomenological relation. According to Eqs.~(\ref{eq1})-(\ref{eq4}), the range of $\cos\delta$ and confidence level of exclusion are calculated and listed in Table~\ref{tab:3cases_NH} (NH) and Table~\ref{tab:3cases_IH} (IH). In the table we also classify consistency between the results and the natural limit by the same labels used in Table~\ref{tab:pairclass}.

\begin{table*}[t]
\caption{\label{tab:3cases_NH}
Three cases of pair combination in NH.}
\begin{ruledtabular}
\begin{tabular}{cccccccccc}
\multirow{2}{*}{Case}&\multirow{2}{*}{Pairs included}&\multirow{2}{*}{$\overline{\cos\delta}$}&\multicolumn{2}{c}{$\sigma_{\overline{\cos\delta}}$}&\multicolumn{3}{c}{Self consistency}&\multicolumn{2}{c}{Natural limit consistency}\\
\cline{4-5}\cline{6-8}\cline{9-10}\\[-.27cm]
&&&Unscaled&Scaled&$\chi^2$&p-value&Exclusion level&Unscaled&Scaled\\
\\[-.27cm] \hline
 trimaximal mixing   &\cirn2~\cirn3~\cirn4&0.257 &0.263&0.420&5.0857&0.079 &1-2 $\sigma$&4 \FiveStarOpen&4 \FiveStarOpen\\[.1cm]
 $\mu$-$\tau$ symmetry&\cirn1~\cirn2~\cirn5&-0.134&0.168&0.179&2.2629&0.323 &$<$1$\sigma$&5 \FiveStarOpen&5 \FiveStarOpen\\[.1cm]
 $\mu$-$\tau$ \& trimaximal&\cirn1~\cirn2~\cirn3~\cirn4~\cirn5&-0.046&0.166&0.225&7.2975&0.121&1-2 $\sigma$&5 \FiveStarOpen&5 \FiveStarOpen\\[.1cm]
\end{tabular}
\end{ruledtabular}
\end{table*}

\begin{table*}[t]
\caption{\label{tab:3cases_IH}
Three cases of pair combination in IH.}
\begin{ruledtabular}
\begin{tabular}{cccccccccc}
\multirow{2}{*}{Case}&\multirow{2}{*}{Pairs included}&\multirow{2}{*}{$\overline{\cos\delta}$}&\multicolumn{2}{c}{$\sigma_{\overline{\cos\delta}}$}&\multicolumn{3}{c}{Self consistency}&\multicolumn{2}{c}{Natural limit consistency}\\
\cline{4-5}\cline{6-8}\cline{9-10}\\[-.27cm]
&&&Unscaled&Scaled&$\chi^2$&p-value&Exclusion level&Unscaled&Scaled\\
\\[-.27cm] \hline
 trimaximal mixing   &\cirn2~\cirn3~\cirn4&-0.657&0.228&0.416&6.6486&0.036 &2-3 $\sigma$&4 \FiveStarOpen&3 \FiveStarOpen\\[.1cm]
 $\mu$-$\tau$ symmetry&\cirn1~\cirn2~\cirn5&0.166&0.155&0.364&11.0783&0.0039&2-3 $\sigma$&5 \FiveStarOpen&4 \FiveStarOpen\\[.1cm]
 $\mu$-$\tau$ \& trimaximal&\cirn1~\cirn2~\cirn3~\cirn4~\cirn5&0.015&0.139&0.327&21.9753&0.0002&3-4 $\sigma$&5 \FiveStarOpen&5 \FiveStarOpen\\[.1cm]
\end{tabular}
\end{ruledtabular}
\end{table*}

The same procedure is performed considering the $\mu$-$\tau$ symmetry and the combination of the two relations. Here some explanation is necessary. Obviously, pair \cirn{\small{5}} has no influence on $\cos\delta$. When combined with other pairs, it simply contributes $\Delta\chi^2=(\theta_{23}-45^\circ)^2/\sigma_{\theta_{23}}^2=0.81~(\textrm{NH})$ or $4.18~(\textrm{IH})$ to $\chi^2$.

From Table~\ref{tab:3cases_NH}, we find that the three cases all give ranges compatible with the maximal {\it CP} violation ($\cos\delta=0$) in $1\sigma$ errors, regardless of whether errors are scaled or not.

On the other hand, Table~\ref{tab:3cases_IH} shows that the $\mu$-$\tau$ symmetry and the trimaximal mixing give ranges compatible with the maximal {\it CP} violation in $2\sigma$~(unscaled)/$1\sigma$~(scaled) and $3\sigma$~(unscaled)/$2\sigma$~(scaled) errors, respectively. The combination of the two relations fits the maximal {\it CP} violation well. However, the three cases in IH all lead to low p-values, indicating a low self-consistency of the combinations in IH.

\subsection{Bipair combination}

In this part we consider all bipair combinations. That is, any two pairs in Sec.~\ref{sec:singlepair} are considered to examine their consistency and constraints on $\delta$.

However, not all of the pairs are suitable for combination: similar to pair \cirn{\small{5}}, \cirn{\small{10}}~\cirn{\small{11}}~\cirn{\small{16}}~\cirn{\small{17}}~\cirn{\small{18}}~\cirn{\small{19}} would simply add to $\chi^2$. Each pair of \cirn{\small{10}}~\cirn{\small{16}}~\cirn{\small{19}} contributes $\Delta\chi^2=24.55~(\textrm{NH})$ or $47.10~(\textrm{IH})$, and each pair of \cirn{\small{11}}~\cirn{\small{17}}~\cirn{\small{18}} contributes $\Delta\chi^2=21.05~(\textrm{NH})$ or $17.24~(\textrm{IH})$. Since we are combining no more than 3 pairs, $\chi^2>17.24$ leads to p-value$<$0.00018 (over 3$\sigma$ level of exclusion). Therefore, we do not include pairs \cirn{\small{10}}~\cirn{\small{16}}~\cirn{\small{19}}~\cirn{\small{11}}~\cirn{\small{17}}~\cirn{\small{18}} in combination.

What is more, pairs \cirn{\small{12}} and \cirn{\small{15}} have the same constraint, as well as \cirn{\small{13}} and \cirn{\small{14}}. Therefore we do not include \cirn{\small{14}}, \cirn{\small{15}} for conciseness. The results are listed in Table~\ref{tab:bipair_NH} (NH) and Table~\ref{tab:bipair_IH} (IH).

\begin{longtable*}[H]{ccccccccccccccccccccccccc}
\caption{\label{tab:bipair_NH}
Bipair combinations in NH.}\\
\hline\\[-3mm]
\hline\\[-.3cm]
\multirow{2}{*}{Pairs included}&&\multirow{2}{*}{$\overline{\cos\delta}$}&&\multicolumn{3}{c}{$\sigma_{\overline{\cos\delta}}$}&&\multicolumn{5}{c}{Self consistency}&&\multicolumn{3}{c}{Natural limit consistency}\\
\cline{5-7}\cline{9-13}\cline{15-17}\\[-.27cm]
&&&&Unscaled&&Scaled&&$\chi^2$&&p-value&&Exclusion level&&Unscaled&&Scaled\\
\\[-.4cm] \hline
\endhead
\hline \multicolumn{18  }{r}{{Continued on next page}} \\ \hline \hline
\endfoot
\hline \hline
\endlastfoot
\cirn{1}~\cirn{2}	&&	-0.134 	&&	0.168 	&&	0.202 	&&	1.4529 	&&	2.28$\TIMES10^{-1}$&&	1-2	$\sigma$&&	5	\FiveStarOpen&&	5	\FiveStarOpen\\	
\cirn{1}~\cirn{3}	&&	-0.064 	&&	0.189 	&&	0.427 	&&	5.1242 	&&	2.36$\TIMES10^{-2}$&&	2-3	$\sigma$&&	5	\FiveStarOpen&&	4	\FiveStarOpen\\	
\cirn{1}~\cirn{4}	&&	-0.213 	&&	0.145 	&&	0.145 	&&	0.0548 	&&	8.15$\TIMES10^{-1}$&&	$<$1$\sigma$&&	5	\FiveStarOpen&&	5	\FiveStarOpen\\	
\cirn{1}~\cirn{5}	&&	-0.201 	&&	0.157 	&&	0.157 	&&	0.8100 	&&	3.68$\TIMES10^{-1}$&&	$<$1$\sigma$&&	5	\FiveStarOpen&&	5	\FiveStarOpen\\	
\cirn{1}~\cirn{6}	&&	0.738 	&&	0.153 	&&	0.875 	&&	32.5828 &&	1.14$\TIMES10^{-8}$&&	$>$5$\sigma$&&	4	\FiveStarOpen&&	3	\FiveStarOpen\\	
\cirn{1}~\cirn{7}	&&	-0.711 	&&	0.087 	&&	0.536 	&&	37.9554 &&	7.24$\TIMES10^{-10}$&&	$>$5$\sigma$&&	5	\FiveStarOpen&&	3	\FiveStarOpen\\	
\cirn{1}~\cirn{8}	&&	-0.230 	&&	0.145 	&&	0.145 	&&	0.9028 	&&	3.42$\TIMES10^{-1}$&&	$<$1$\sigma$&&	5	\FiveStarOpen&&	5	\FiveStarOpen\\	
\cirn{1}~\cirn{9}	&&	-0.143 	&&	0.173 	&&	0.441 	&&	6.5201 	&&	1.07$\TIMES10^{-2}$&&	2-3	$\sigma$&&	5	\FiveStarOpen&&	4	\FiveStarOpen\\	
\cirn{1}~\cirn{12}	&&	-0.554 	&&	0.109 	&&	0.771 	&&	49.8493 &&	1.66$\TIMES10^{-12}$&&	$>$5$\sigma$&&	5	\FiveStarOpen&&	3	\FiveStarOpen\\	
\cirn{1}~\cirn{13}	&&	1.301 	&&	0.115 	&&	0.896 	&&	60.5831 &&	7.11$\TIMES10^{-15}$&&	$>$5$\sigma$&&	1	\FiveStarOpen&&	2	\FiveStarOpen\\	
\cirn{2}~\cirn{3}	&&	0.647 	&&	0.286 	&&	0.324 	&&	1.2847 	&&	2.57$\TIMES10^{-1}$&&	1-2	$\sigma$&&	4	\FiveStarOpen&&	4	\FiveStarOpen\\	
\cirn{2}~\cirn{4}	&&	-0.014 	&&	0.321 	&&	0.383 	&&	1.4264 	&&	2.32$\TIMES10^{-1}$&&	1-2	$\sigma$&&	5	\FiveStarOpen&&	4	\FiveStarOpen\\	
\cirn{2}~\cirn{5}	&&	0.479 	&&	0.373 	&&	0.373 	&&	0.8100 	&&	3.68$\TIMES10^{-1}$&&	$<$1$\sigma$&&	4	\FiveStarOpen&&	4	\FiveStarOpen\\	
\cirn{2}~\cirn{6}	&&	1.178 	&&	0.169 	&&	0.512 	&&	9.1718 	&&	2.46$\TIMES10^{-3}$&&	3-4	$\sigma$&&	1	\FiveStarOpen&&	2	\FiveStarOpen\\	
\cirn{2}~\cirn{7}	&&	-1.176 	&&	0.127 	&&	0.404 	&&	10.1180 &&	1.47$\TIMES10^{-3}$&&	3-4	$\sigma$&&	1	\FiveStarOpen&&	2	\FiveStarOpen\\	
\cirn{2}~\cirn{8}	&&	-0.023 	&&	0.417 	&&	0.655 	&&	2.4671 	&&	1.16$\TIMES10^{-1}$&&	1-2	$\sigma$&&	4	\FiveStarOpen&&	4	\FiveStarOpen\\	
\cirn{2}~\cirn{9}	&&	0.635 	&&	0.316 	&&	0.632 	&&	4.0004 	&&	4.55$\TIMES10^{-2}$&&	2-3	$\sigma$&&	4	\FiveStarOpen&&	3	\FiveStarOpen\\	
\cirn{2}~\cirn{12}	&&	-1.712 	&&	0.236 	&&	1.073 	&&	20.7529 &&	5.23$\TIMES10^{-6}$&&	4-5	$\sigma$&&	0	\FiveStarOpen&&	2	\FiveStarOpen\\	
\cirn{2}~\cirn{13}	&&	1.591 	&&	0.121 	&&	0.522 	&&	18.4531 &&	1.74$\TIMES10^{-5}$&&	4-5	$\sigma$&&	0	\FiveStarOpen&&	1	\FiveStarOpen\\	
\cirn{3}~\cirn{4}	&&	0.134 	&&	0.328 	&&	0.711 	&&	4.6908 	&&	3.03$\TIMES10^{-2}$&&	2-3	$\sigma$&&	4	\FiveStarOpen&&	4	\FiveStarOpen\\	
\cirn{3}~\cirn{5}	&&	1.269 	&&	0.490 	&&	0.490 	&&	0.8100 	&&	3.68$\TIMES10^{-1}$&&	$<$1$\sigma$&&	2	\FiveStarOpen&&	2	\FiveStarOpen\\	
\cirn{3}~\cirn{6}	&&	1.511 	&&	0.168 	&&	0.168 	&&	0.3544 	&&	5.52$\TIMES10^{-1}$&&	$<$1$\sigma$&&	0	\FiveStarOpen&&	0	\FiveStarOpen\\	
\cirn{3}~\cirn{7}	&&	-1.168 	&&	0.127 	&&	0.510 	&&	16.2066 &&	5.68$\TIMES10^{-5}$&&	4-5	$\sigma$&&	1	\FiveStarOpen&&	2	\FiveStarOpen\\	
\cirn{3}~\cirn{8}	&&	0.306 	&&	0.459 	&&	1.068 	&&	5.4052 	&&	2.01$\TIMES10^{-2}$&&	2-3	$\sigma$&&	4	\FiveStarOpen&&	3	\FiveStarOpen\\	
\cirn{3}~\cirn{9}	&&	1.469 	&&	0.425 	&&	0.588 	&&	1.9126 	&&	1.67$\TIMES10^{-1}$&&	1-2	$\sigma$&&	1	\FiveStarOpen&&	2	\FiveStarOpen\\	
\cirn{3}~\cirn{12}	&&	-1.703 	&&	0.242 	&&	1.261 	&&	27.2497 &&	1.79$\TIMES10^{-7}$&&	$>$5$\sigma$&&	1	\FiveStarOpen&&	2	\FiveStarOpen\\	
\cirn{3}~\cirn{13}	&&	1.768 	&&	0.141 	&&	0.184 	&&	1.7137 	&&	1.91$\TIMES10^{-1}$&&	1-2	$\sigma$&&	0	\FiveStarOpen&&	0	\FiveStarOpen\\	
\cirn{4}~\cirn{5}	&&	-0.311 	&&	0.466 	&&	0.466 	&&	0.8100 	&&	3.68$\TIMES10^{-1}$&&	$<$1$\sigma$&&	4	\FiveStarOpen&&	4	\FiveStarOpen\\	
\cirn{4}~\cirn{6}	&&	1.131 	&&	0.184 	&&	0.780 	&&	17.9223 &&	2.30$\TIMES10^{-5}$&&	4-5	$\sigma$&&	2	\FiveStarOpen&&	2	\FiveStarOpen\\	
\cirn{4}~\cirn{7}	&&	-1.224 	&&	0.134 	&&	0.215 	&&	2.5580 	&&	1.10$\TIMES10^{-1}$&&	1-2	$\sigma$&&	1	\FiveStarOpen&&	1	\FiveStarOpen\\	
\cirn{4}~\cirn{8}	&&	-0.479 	&&	0.420 	&&	0.420 	&&	0.3792 	&&	5.38$\TIMES10^{-1}$&&	$<$1$\sigma$&&	4	\FiveStarOpen&&	4	\FiveStarOpen\\	
\cirn{4}~\cirn{9}	&&	-0.010 	&&	0.387 	&&	0.984 	&&	6.4635 	&&	1.10$\TIMES10^{-2}$&&	2-3	$\sigma$&&	4	\FiveStarOpen&&	4	\FiveStarOpen\\	
\cirn{4}~\cirn{12}	&&	-1.916 	&&	0.239 	&&	0.718 	&&	8.9971 	&&	2.70$\TIMES10^{-3}$&&	2-3	$\sigma$&&	0	\FiveStarOpen&&	1	\FiveStarOpen\\	
\cirn{4}~\cirn{13}	&&	1.606 	&&	0.127 	&&	0.664 	&&	27.4527 &&	1.61$\TIMES10^{-7}$&&	$>$5$\sigma$&&	0	\FiveStarOpen&&	2	\FiveStarOpen\\	
\cirn{5}~\cirn{6}	&&	1.553 	&&	0.175 	&&	0.175 	&&	0.8100 	&&	3.68$\TIMES10^{-1}$&&	$<$1$\sigma$&&	0	\FiveStarOpen&&	0	\FiveStarOpen\\	
\cirn{5}~\cirn{7}	&&	-1.275 	&&	0.146 	&&	0.146 	&&	0.8100 	&&	3.68$\TIMES10^{-1}$&&	$<$1$\sigma$&&	1	\FiveStarOpen&&	1	\FiveStarOpen\\	
\cirn{5}~\cirn{8}	&&	-0.878 	&&	0.893 	&&	0.893 	&&	0.8100 	&&	3.68$\TIMES10^{-1}$&&	$<$1$\sigma$&&	3	\FiveStarOpen&&	3	\FiveStarOpen\\	
\cirn{5}~\cirn{9}	&&	3.196 	&&	0.922 	&&	0.922 	&&	0.8100 	&&	3.68$\TIMES10^{-1}$&&	$<$1$\sigma$&&	1	\FiveStarOpen&&	1	\FiveStarOpen\\	
\cirn{5}~\cirn{12}	&&	-2.238 	&&	0.202 	&&	0.202 	&&	0.8100 	&&	3.68$\TIMES10^{-1}$&&	$<$1$\sigma$&&	0	\FiveStarOpen&&	0	\FiveStarOpen\\	
\cirn{5}~\cirn{13}	&&	1.836 	&&	0.167 	&&	0.167 	&&	0.8100 	&&	3.68$\TIMES10^{-1}$&&	$<$1$\sigma$&&	0	\FiveStarOpen&&	0	\FiveStarOpen\\	
\cirn{6}~\cirn{7}	&&	-0.521 	&&	0.108 	&&	1.250 	&&	133.6315&&	0	               &&	$>$5$\sigma$&&	5	\FiveStarOpen&&	3	\FiveStarOpen\\	
\cirn{6}~\cirn{8}	&&	1.346 	&&	0.200 	&&	0.678 	&&	11.4907 &&	6.99$\TIMES10^{-4}$&&	3-4	$\sigma$&&	1	\FiveStarOpen&&	2	\FiveStarOpen\\	
\cirn{6}~\cirn{9}	&&	1.579 	&&	0.169 	&&	0.209 	&&	1.5269 	&&	2.17$\TIMES10^{-1}$&&	1-2	$\sigma$&&	0	\FiveStarOpen&&	1	\FiveStarOpen\\	
\cirn{6}~\cirn{12}	&&	0.077 	&&	0.164 	&&	1.848 	&&	127.5130&&	0                  &&	$>$5$\sigma$&&	5	\FiveStarOpen&&	3	\FiveStarOpen\\	
\cirn{6}~\cirn{13}	&&	1.708 	&&	0.101 	&&	0.141 	&&	1.9665 	&&	1.61$\TIMES10^{-1}$&&	1-2	$\sigma$&&	0	\FiveStarOpen&&	0	\FiveStarOpen\\	
\cirn{7}~\cirn{8}	&&	-1.267 	&&	0.143 	&&	0.143 	&&	0.1509 	&&	6.98$\TIMES10^{-1}$&&	$<$1$\sigma$&&	1	\FiveStarOpen&&	1	\FiveStarOpen\\	
\cirn{7}~\cirn{9}	&&	-1.226 	&&	0.138 	&&	0.464 	&&	11.3660 &&	7.48$\TIMES10^{-4}$&&	3-4	$\sigma$&&	1	\FiveStarOpen&&	2	\FiveStarOpen\\	
\cirn{7}~\cirn{12}	&&	-1.565 	&&	0.144 	&&	0.442 	&&	9.4236 	&&	2.14$\TIMES10^{-3}$&&	3-4	$\sigma$&&	0	\FiveStarOpen&&	1	\FiveStarOpen\\	
\cirn{7}~\cirn{13}	&&	0.188 	&&	0.092 	&&	1.552 	&&	285.1556&&	0	               &&	$>$5$\sigma$&&	5	\FiveStarOpen&&	3	\FiveStarOpen\\	
\cirn{8}~\cirn{9}	&&	-0.011 	&&	0.609 	&&	1.668 	&&	7.5110 	&&	6.13$\TIMES10^{-3}$&&	2-3	$\sigma$&&	4	\FiveStarOpen&&	3	\FiveStarOpen\\	
\cirn{8}~\cirn{12}	&&	-2.202 	&&	0.207 	&&	0.217 	&&	1.1014 	&&	2.94$\TIMES10^{-1}$&&	1-2	$\sigma$&&	0	\FiveStarOpen&&	0	\FiveStarOpen\\	
\cirn{8}~\cirn{13}	&&	1.727 	&&	0.137 	&&	0.533 	&&	15.0276 &&	1.06$\TIMES10^{-4}$&&	3-4	$\sigma$&&	0	\FiveStarOpen&&	1	\FiveStarOpen\\	
\cirn{9}~\cirn{12}	&&	-2.060 	&&	0.239 	&&	0.968 	&&	16.4199 &&	5.07$\TIMES10^{-5}$&&	4-5	$\sigma$&&	0	\FiveStarOpen&&	1	\FiveStarOpen\\	
\cirn{9}~\cirn{13}	&&	1.859 	&&	0.171 	&&	0.175 	&&	1.0452 	&&	3.07$\TIMES10^{-1}$&&	1-2	$\sigma$&&	0	\FiveStarOpen&&	0	\FiveStarOpen\\	
\cirn{12}~\cirn{13}	&&	0.992 	&&	0.119 	&&	1.651 	&&	191.2192&&	0	               &&	$>$5$\sigma$&&	3	\FiveStarOpen&&	3	\FiveStarOpen\\	
\end{longtable*}

\begin{longtable*}[H]{ccccccccccccccccccccccccc}
\caption{\label{tab:bipair_IH}
Bipair combinations in IH.}\\
\hline\\[-3mm]
\hline\\[-.3cm]
\multirow{2}{*}{Pairs included}&&\multirow{2}{*}{$\overline{\cos\delta}$}&&\multicolumn{3}{c}{$\sigma_{\overline{\cos\delta}}$}&&\multicolumn{5}{c}{Self consistency}&&\multicolumn{3}{c}{Natural limit consistency}\\
\cline{5-7}\cline{9-13}\cline{15-17}\\[-.27cm]
&&&&Unscaled&&Scaled&&$\chi^2$&&p-value&&Exclusion level&&Unscaled&&Scaled\\
\\[-.4cm] \hline
\endhead
\hline \multicolumn{18  }{r}{{Continued on next page}} \\ \hline \hline
\endfoot
\hline \hline
\endlastfoot
\cirn{1}~\cirn{2}	&&	0.166 	&&	0.155 	&&	0.406 	&&	6.8944 	&&	8.65$\TIMES10^{-3}$&&	     2-3	$\sigma$&&	5	\FiveStarOpen&&	4	\FiveStarOpen\\
\cirn{1}~\cirn{3}	&&	0.305 	&&	0.136 	&&	0.136 	&&	0.5355 	&&	4.64$\TIMES10^{-1}$&&	     $<$1	$\sigma$&&	5	\FiveStarOpen&&	5	\FiveStarOpen\\
\cirn{1}~\cirn{4}	&&	0.149 	&&	0.159 	&&	0.573 	&&	12.9107 &&	3.27$\TIMES10^{-4}$&&	     3-4	$\sigma$&&	5	\FiveStarOpen&&	4	\FiveStarOpen\\
\cirn{1}~\cirn{5}	&&	0.336 	&&	0.136 	&&	0.279 	&&	4.1839 	&&	4.08$\TIMES10^{-2}$&&	     2-3	$\sigma$&&	5	\FiveStarOpen&&	4	\FiveStarOpen\\
\cirn{1}~\cirn{6}	&&	0.759 	&&	0.082 	&&	0.427 	&&	27.3863 &&	1.67$\TIMES10^{-7}$&&	     $>$5	$\sigma$&&	4	\FiveStarOpen&&	3	\FiveStarOpen\\
\cirn{1}~\cirn{7}	&&	-0.552 	&&	0.125 	&&	0.990 	&&	62.9428 &&	2.11$\TIMES10^{-15}$&&	     $>$5	$\sigma$&&	5	\FiveStarOpen&&	3	\FiveStarOpen\\
\cirn{1}~\cirn{8}	&&	0.233 	&&	0.154 	&&	0.661 	&&	18.4609 &&	1.73$\TIMES10^{-5}$&&	     4-5	$\sigma$&&	5	\FiveStarOpen&&	4	\FiveStarOpen\\
\cirn{1}~\cirn{9}	&&	0.330 	&&	0.136 	&&	0.136 	&&	0.0751 	&&	7.84$\TIMES10^{-1}$&&	     $<$1	$\sigma$&&	5	\FiveStarOpen&&	5	\FiveStarOpen\\
\cirn{1}~\cirn{12}	&&	-1.009 	&&	0.098 	&&	0.974 	&&	98.0825 &&	0	               &&	     $>$5	$\sigma$&&	2	\FiveStarOpen&&	2	\FiveStarOpen\\
\cirn{1}~\cirn{13}	&&	0.786 	&&	0.102 	&&	0.848 	&&	69.8745 &&	1.11$\TIMES10^{-16}$&&	     $>$5	$\sigma$&&	4	\FiveStarOpen&&	3	\FiveStarOpen\\
\cirn{2}~\cirn{3}	&&	-0.373 	&&	0.272 	&&	0.397 	&&	2.1345 	&&	1.44$\TIMES10^{-1}$&&	     1-2	$\sigma$&&	4	\FiveStarOpen&&	4	\FiveStarOpen\\
\cirn{2}~\cirn{4}	&&	-0.994 	&&	0.251 	&&	0.341 	&&	1.8378 	&&	1.75$\TIMES10^{-1}$&&	     1-2	$\sigma$&&	3	\FiveStarOpen&&	3	\FiveStarOpen\\
\cirn{2}~\cirn{5}	&&	-0.801 	&&	0.325 	&&	0.664 	&&	4.1839 	&&	4.08$\TIMES10^{-2}$&&	     2-3	$\sigma$&&	3	\FiveStarOpen&&	3	\FiveStarOpen\\
\cirn{2}~\cirn{6}	&&	1.036 	&&	0.111 	&&	0.534 	&&	22.9259 &&	1.68$\TIMES10^{-6}$&&	     4-5	$\sigma$&&	2	\FiveStarOpen&&	2	\FiveStarOpen\\
\cirn{2}~\cirn{7}	&&	-1.384 	&&	0.154 	&&	0.398 	&&	6.6574 	&&	9.87$\TIMES10^{-3}$&&	     2-3	$\sigma$&&	1	\FiveStarOpen&&	2	\FiveStarOpen\\
\cirn{2}~\cirn{8}	&&	-1.036 	&&	0.270 	&&	0.833 	&&	9.5159 	&&	2.04$\TIMES10^{-3}$&&	     3-4	$\sigma$&&	2	\FiveStarOpen&&	2	\FiveStarOpen\\
\cirn{2}~\cirn{9}	&&	-0.581 	&&	0.319 	&&	0.391 	&&	1.5035 	&&	2.20$\TIMES10^{-1}$&&	     1-2	$\sigma$&&	4	\FiveStarOpen&&	4	\FiveStarOpen\\
\cirn{2}~\cirn{12}	&&	-1.564 	&&	0.111 	&&	0.339 	&&	9.3117 	&&	2.28$\TIMES10^{-3}$&&	     3-4	$\sigma$&&	0	\FiveStarOpen&&	1	\FiveStarOpen\\
\cirn{2}~\cirn{13}	&&	1.662 	&&	0.191 	&&	1.336 	&&	49.1950 &&	2.32$\TIMES10^{-12}$&&	     $>$5	$\sigma$&&	0	\FiveStarOpen&&	2	\FiveStarOpen\\
\cirn{3}~\cirn{4}	&&	-0.553 	&&	0.300 	&&	0.756 	&&	6.3592 	&&	1.17$\TIMES10^{-2}$&&	     2-3	$\sigma$&&	4	\FiveStarOpen&&	3	\FiveStarOpen\\
\cirn{3}~\cirn{5}	&&	-0.006 	&&	0.411 	&&	0.840 	&&	4.1839 	&&	4.08$\TIMES10^{-2}$&&	     2-3	$\sigma$&&	4	\FiveStarOpen&&	4	\FiveStarOpen\\
\cirn{3}~\cirn{6}	&&	1.116 	&&	0.115 	&&	0.289 	&&	6.3013 	&&	1.21$\TIMES10^{-2}$&&	     2-3	$\sigma$&&	1	\FiveStarOpen&&	2	\FiveStarOpen\\
\cirn{3}~\cirn{7}	&&	-1.321 	&&	0.167 	&&	0.664 	&&	15.8375 &&	6.90$\TIMES10^{-5}$&&	     3-4	$\sigma$&&	1	\FiveStarOpen&&	2	\FiveStarOpen\\
\cirn{3}~\cirn{8}	&&	-0.490 	&&	0.347 	&&	1.302 	&&	14.0912 &&	1.74$\TIMES10^{-4}$&&	     3-4	$\sigma$&&	4	\FiveStarOpen&&	3	\FiveStarOpen\\
\cirn{3}~\cirn{9}	&&	0.022 	&&	0.363 	&&	0.363 	&&	0.0185 	&&	8.92$\TIMES10^{-1}$&&	     $<$1	$\sigma$&&	4	\FiveStarOpen&&	4	\FiveStarOpen\\
\cirn{3}~\cirn{12}	&&	-1.548 	&&	0.115 	&&	0.506 	&&	19.2342 &&	1.16$\TIMES10^{-5}$&&	     4-5	$\sigma$&&	0	\FiveStarOpen&&	1	\FiveStarOpen\\
\cirn{3}~\cirn{13}	&&	1.954 	&&	0.196 	&&	0.921 	&&	22.0236 &&	2.69$\TIMES10^{-6}$&&	     4-5	$\sigma$&&	0	\FiveStarOpen&&	1	\FiveStarOpen\\
\cirn{4}~\cirn{5}	&&	-1.597 	&&	0.453 	&&	0.927 	&&	4.1839 	&&	4.08$\TIMES10^{-2}$&&	     2-3	$\sigma$&&	1	\FiveStarOpen&&	2	\FiveStarOpen\\
\cirn{4}~\cirn{6}	&&	1.054 	&&	0.113 	&&	0.602 	&&	28.2693 &&	1.06$\TIMES10^{-7}$&&	     $>$5	$\sigma$&&	2	\FiveStarOpen&&	2	\FiveStarOpen\\
\cirn{4}~\cirn{7}	&&	-1.648 	&&	0.160 	&&	0.160 	&&	0.0148 	&&	9.03$\TIMES10^{-1}$&&	     $<$1	$\sigma$&&	0	\FiveStarOpen&&	0	\FiveStarOpen\\
\cirn{4}~\cirn{8}	&&	-1.949 	&&	0.381 	&&	0.847 	&&	4.9330 	&&	2.63$\TIMES10^{-2}$&&	     2-3	$\sigma$&&	1	\FiveStarOpen&&	1	\FiveStarOpen\\
\cirn{4}~\cirn{9}	&&	-0.942 	&&	0.401 	&&	0.831 	&&	4.2837 	&&	3.85$\TIMES10^{-2}$&&	     2-3	$\sigma$&&	3	\FiveStarOpen&&	3	\FiveStarOpen\\
\cirn{4}~\cirn{12}	&&	-1.704 	&&	0.131 	&&	0.131 	&&	0.0638 	&&	8.01$\TIMES10^{-1}$&&	     $<$1	$\sigma$&&	0	\FiveStarOpen&&	0	\FiveStarOpen\\
\cirn{4}~\cirn{13}	&&	1.780 	&&	0.200 	&&	1.431 	&&	51.4547 &&	7.33$\TIMES10^{-13}$&&	     $>$5	$\sigma$&&	0	\FiveStarOpen&&	2	\FiveStarOpen\\
\cirn{5}~\cirn{6}	&&	1.191 	&&	0.125 	&&	0.255 	&&	4.1839 	&&	4.08$\TIMES10^{-2}$&&	     2-3	$\sigma$&&	1	\FiveStarOpen&&	2	\FiveStarOpen\\
\cirn{5}~\cirn{7}	&&	-1.656 	&&	0.170 	&&	0.348 	&&	4.1839 	&&	4.08$\TIMES10^{-2}$&&	     2-3	$\sigma$&&	0	\FiveStarOpen&&	1	\FiveStarOpen\\
\cirn{5}~\cirn{8}	&&	-3.989 	&&	0.810 	&&	1.657 	&&	4.1839 	&&	4.08$\TIMES10^{-2}$&&	     2-3	$\sigma$&&	0	\FiveStarOpen&&	1	\FiveStarOpen\\
\cirn{5}~\cirn{9}	&&	0.111 	&&	0.769 	&&	1.572 	&&	4.1839 	&&	4.08$\TIMES10^{-2}$&&	     2-3	$\sigma$&&	4	\FiveStarOpen&&	3	\FiveStarOpen\\
\cirn{5}~\cirn{12}	&&	-1.714 	&&	0.139 	&&	0.283 	&&	4.1839 	&&	4.08$\TIMES10^{-2}$&&	     2-3	$\sigma$&&	0	\FiveStarOpen&&	1	\FiveStarOpen\\
\cirn{5}~\cirn{13}	&&	2.386 	&&	0.189 	&&	0.387 	&&	4.1839 	&&	4.08$\TIMES10^{-2}$&&	     2-3	$\sigma$&&	0	\FiveStarOpen&&	0	\FiveStarOpen\\
\cirn{6}~\cirn{7}	&&	0.399 	&&	0.099 	&&	1.275 	&&	167.4270&&	0	               &&	     $>$5	$\sigma$&&	5	\FiveStarOpen&&	3	\FiveStarOpen\\
\cirn{6}~\cirn{8}	&&	1.117 	&&	0.118 	&&	0.612 	&&	26.7374 &&	2.33$\TIMES10^{-7}$&&	     $>$5	$\sigma$&&	2	\FiveStarOpen&&	2	\FiveStarOpen\\
\cirn{6}~\cirn{9}	&&	1.172 	&&	0.122 	&&	0.139 	&&	1.2834 	&&	2.57$\TIMES10^{-1}$&&	     1-2	$\sigma$&&	1	\FiveStarOpen&&	1	\FiveStarOpen\\
\cirn{6}~\cirn{12}	&&	-0.196 	&&	0.084 	&&	1.451 	&&	299.1174&&	0	               &&	     $>$5	$\sigma$&&	5	\FiveStarOpen&&	3	\FiveStarOpen\\
\cirn{6}~\cirn{13}	&&	1.523 	&&	0.114 	&&	0.536 	&&	21.9778 &&	2.76$\TIMES10^{-6}$&&	     4-5	$\sigma$&&	0	\FiveStarOpen&&	2	\FiveStarOpen\\
\cirn{7}~\cirn{8}	&&	-1.718 	&&	0.163 	&&	0.376 	&&	5.3578 	&&	2.06$\TIMES10^{-2}$&&	     2-3	$\sigma$&&	0	\FiveStarOpen&&	1	\FiveStarOpen\\
\cirn{7}~\cirn{9}	&&	-1.527 	&&	0.175 	&&	0.459 	&&	6.8914 	&&	8.66$\TIMES10^{-3}$&&	     2-3	$\sigma$&&	0	\FiveStarOpen&&	1	\FiveStarOpen\\
\cirn{7}~\cirn{12}	&&	-1.691 	&&	0.105 	&&	0.105 	&&	0.0736 	&&	7.86$\TIMES10^{-1}$&&	     $<$1	$\sigma$&&	0	\FiveStarOpen&&	0	\FiveStarOpen\\
\cirn{7}~\cirn{13}	&&	0.068 	&&	0.142 	&&	1.999 	&&	199.5146&&	0	               &&	     $>$5	$\sigma$&&	5	\FiveStarOpen&&	3	\FiveStarOpen\\
\cirn{8}~\cirn{9}	&&	-1.111 	&&	0.543 	&&	1.875 	&&	11.9311 &&	5.52$\TIMES10^{-4}$&&	     3-4	$\sigma$&&	2	\FiveStarOpen&&	2	\FiveStarOpen\\
\cirn{8}~\cirn{12}	&&	-1.762 	&&	0.144 	&&	0.325 	&&	5.1237 	&&	2.36$\TIMES10^{-2}$&&	     2-3	$\sigma$&&	0	\FiveStarOpen&&	1	\FiveStarOpen\\
\cirn{8}~\cirn{13}	&&	2.097 	&&	0.212 	&&	1.326 	&&	39.2286 &&	3.77$\TIMES10^{-10}$&&	     $>$5	$\sigma$&&	0	\FiveStarOpen&&	2	\FiveStarOpen\\
\cirn{9}~\cirn{12}	&&	-1.645 	&&	0.126 	&&	0.349 	&&	7.6307 	&&	5.74$\TIMES10^{-3}$&&	     2-3	$\sigma$&&	0	\FiveStarOpen&&	1	\FiveStarOpen\\
\cirn{9}~\cirn{13}	&&	2.287 	&&	0.197 	&&	0.465 	&&	5.5475 	&&	1.85$\TIMES10^{-2}$&&	     2-3	$\sigma$&&	0	\FiveStarOpen&&	1	\FiveStarOpen\\
\cirn{12}~\cirn{13}	&&	-0.734 	&&	0.106 	&&	1.748 	&&	272.4518&&	0	               &&	     $>$5	$\sigma$&&	4	\FiveStarOpen&&	3	\FiveStarOpen\\
\end{longtable*}

It is worthwhile to point out that in the search for new mixing patterns, two pairs may automatically imply a third one. For example, pairs \cirn{\small{2}}~\cirn{\small{3}} contain the same matrix element $U_{22}$, and it leads to another pair $|U_{12}|=|U_{32}|$, which is \cirn{\small{4}}. Another example is that any two of the pairs in the $\mu$-$\tau$ symmetry would lead to the third one due to the unitarity of the mixing matrix. However, the combined results of \cirn{\small{2}}~\cirn{\small{3}} are different from that of \cirn{\small{4}}, according to Tables~\ref{tab:bipair_NH},~\ref{tab:bipair_IH} and Fig.~\ref{fig:cosdelta}.

The reason is that the implication of a third pair is based on the condition that the two pairs hold precisely, i.e., the modulus of elements of each pair are precisely equal to each other. But for experimental results, pairs hold with errors. Therefore, pairs \cirn{\small{2}}~\cirn{\small{3}} do not necessarily imply the third pair \cirn{\small{4}}. Another example can be seen from Eq.~(\ref{eq5}): while pairs \cirn{\small{6}}~\cirn{\small{8}} both hold in 3$\sigma$ errors, the corresponding ``third pair,'' i.e., $|U_{21}|=|U_{23}|$ does not hold in 3$\sigma$ error. Therefore we do not consider correlations among pairs in our discussion due to the presence of experimental errors.

While a total number of 55 cases are listed in Tables~\ref{tab:bipair_NH}~and~\ref{tab:bipair_IH}, not all of them are self-consistent. When regarding $3\sigma$ as a dividing line of self-consistency (i.e., regarding p-value$>$0.0027 as self-consistent), there are 39 cases in NH and 33 cases in IH that are self-consistent. The overall self-consistency in NH exceeds that in IH slightly.

Moreover, many cases give constraints on $\cos\delta$ that are not consistent with natural limit $\cos\delta\in[-1,1]$. Number of cases with central value $\overline{\cos\delta}\in[-1,1]$---namely cases over 3\FiveStarOpen~level---are less than half of the total. The number in NH is 25, and in IH is 23.

The numbers of cases both self-consistent and consistent with natural limit are 18 in NH and 13 in IH. If we have a close look at these cases, we would find them consistent with the maximal {\it CP} violation in $3\sigma$ error range, except for \cirn{\small{2}}~\cirn{\small{4}} in IH and with an unscaled error. Especially, within all 18 cases in NH, 11 of them are compatible with the maximal {\it CP} violation within $1\sigma$ range, whether the errors are scaled or not. The detailed results are listed in Table~\ref{tab:CPV_NH} (NH) and Table~\ref{tab:CPV_IH} (IH) in Sec.~\ref{sec:mCPV}.

\subsection{Tripair combination}

Similarly we consider all tripair combinations among \cirn{\small{1}}-\cirn{\small{9}} \& \cirn{\small{12}}-\cirn{\small{15}}. For there are too many combination cases (286 in total) but most of them are not self-consistent (over $3\sigma$ exclusion level), we do not list combinations over $3\sigma$ exclusion in both NH and IH. The results are listed in Table~\ref{tab:tripair_NH} (NH) and Table~\ref{tab:tripair_IH} (IH).

\begin{longtable*}{cccccccccccccccccc}
\caption{\label{tab:tripair_NH}
Tripair combinations in NH.}\\
\hline\\[-3mm]
\hline\\[-.3cm]
\multirow{2}{*}{Pairs included}&&\multirow{2}{*}{$\overline{\cos\delta}$}&&\multicolumn{3}{c}{$\sigma_{\overline{\cos\delta}}$}&&\multicolumn{5}{c}{Self consistency}&&\multicolumn{3}{c}{Natural limit consistency}\\
\cline{5-7}\cline{9-13}\cline{15-17}\\[-.27cm]
&&&&Unscaled&&Scaled&&$\chi^2$&&p-value&&Exclusion level&&Unscaled&&Scaled\\
\\[-.4cm] \hline
\endhead
\hline \multicolumn{18  }{r}{{Continued on next page}} \\ \hline \hline
\endfoot
\hline \hline
\endlastfoot
 	\cirn{1}~\cirn{2}~\cirn{3}&&  0.046&& 0.195&& 0.333&&	5.8551  &&	5.35$\TIMES10^{-2}$&&	1-2 $\sigma$&&	5   \FiveStarOpen&&	4   \FiveStarOpen\\[.02cm]
 	\cirn{1}~\cirn{2}~\cirn{4}&&	-0.160&&0.151&&	0.151&&	1.6025 	&&	4.49$\TIMES10^{-1}$&&	$<$1$\sigma$&&	5	\FiveStarOpen&&	5	\FiveStarOpen\\[.02cm]
 	\cirn{1}~\cirn{2}~\cirn{5}&&	-0.134&&0.168&&	0.179&&	2.2629 	&&	3.23$\TIMES10^{-1}$&&	$<$1$\sigma$&&	5	\FiveStarOpen&&	5	\FiveStarOpen\\[.02cm]
 	\cirn{1}~\cirn{2}~\cirn{8}&&	-0.179&&0.152&&	0.172&&	2.5660 	&&	2.77$\TIMES10^{-1}$&&	1-2	$\sigma$&&	5	\FiveStarOpen&&	5	\FiveStarOpen\\[.02cm]
 	\cirn{1}~\cirn{2}~\cirn{9}&&	-0.047&&0.188&&	0.366&&	7.5949 	&&	2.24$\TIMES10^{-2}$&&	2-3	$\sigma$&&	5	\FiveStarOpen&&	4	\FiveStarOpen\\[.02cm]
 	\cirn{1}~\cirn{3}~\cirn{4}&&	-0.116&&0.162&&	0.267&&	5.4425 	&&	6.58$\TIMES10^{-2}$&&	1-2	$\sigma$&&	5	\FiveStarOpen&&	5	\FiveStarOpen\\[.02cm]
 	\cirn{1}~\cirn{3}~\cirn{5}&&	-0.064&&0.189&&	0.325&&	5.9342 	&&	5.15$\TIMES10^{-2}$&&	1-2	$\sigma$&&	5	\FiveStarOpen&&	4	\FiveStarOpen\\[.02cm]
 	\cirn{1}~\cirn{3}~\cirn{8}&&	-0.136&&0.165&&	0.296&&	6.4678 	&&	3.94$\TIMES10^{-2}$&&	2-3	$\sigma$&&	5	\FiveStarOpen&&	4	\FiveStarOpen\\[.02cm]
 	\cirn{1}~\cirn{3}~\cirn{9}&&	0.051&& 0.209&&	0.485&&	10.8220 &&	4.47$\TIMES10^{-3}$&&	2-3	$\sigma$&&	5	\FiveStarOpen&&	4	\FiveStarOpen\\[.02cm]
 	\cirn{1}~\cirn{4}~\cirn{5}&&	-0.213&&0.145&&	0.145&&	0.8648 	&&	6.49$\TIMES10^{-1}$&&	$<$1$\sigma$&&	5	\FiveStarOpen&&	5	\FiveStarOpen\\[.02cm]
 	\cirn{1}~\cirn{4}~\cirn{8}&&	-0.237&&0.136&&	0.136&&	0.9291 	&&	6.28$\TIMES10^{-1}$&&	$<$1$\sigma$&&	5	\FiveStarOpen&&	5	\FiveStarOpen\\[.02cm]
 	\cirn{1}~\cirn{4}~\cirn{9}&&	-0.168&&0.155&&	0.282&&	6.6526 	&&	3.59$\TIMES10^{-2}$&&	2-3	$\sigma$&&	5	\FiveStarOpen&&	4	\FiveStarOpen\\[.02cm]
 	\cirn{1}~\cirn{5}~\cirn{8}&&	-0.230&&0.144&&	0.144&&	1.7128 	&&	4.25$\TIMES10^{-1}$&&	$<$1$\sigma$&&	5	\FiveStarOpen&&	5	\FiveStarOpen\\[.02cm]
 	\cirn{1}~\cirn{5}~\cirn{9}&&	-0.143&&0.173&&	0.331&&	7.3301 	&&	2.56$\TIMES10^{-2}$&&	2-3	$\sigma$&&	5	\FiveStarOpen&&	4	\FiveStarOpen\\[.02cm]
 	\cirn{1}~\cirn{8}~\cirn{9}&&	-0.189&&0.155&&	0.302&&	7.6024 	&&	2.23$\TIMES10^{-2}$&&	2-3	$\sigma$&&	5	\FiveStarOpen&&	4	\FiveStarOpen\\[.02cm]
 	\cirn{2}~\cirn{3}~\cirn{4}&&	0.257&& 0.263&&	0.420&&	5.0857 	&&	7.86$\TIMES10^{-2}$&&	1-2	$\sigma$&&	4	\FiveStarOpen&&	4	\FiveStarOpen\\[.02cm]
 	\cirn{2}~\cirn{3}~\cirn{5}&&	0.647&& 0.285&&	0.292&&	2.0947 	&&	3.51$\TIMES10^{-1}$&&	$<$1$\sigma$&&	4	\FiveStarOpen&&	4	\FiveStarOpen\\[.02cm]
 	\cirn{2}~\cirn{3}~\cirn{6}&&	1.187&&	0.160&&	0.344&&	9.2009  &&	1.00$\TIMES10^{-2}$&&	2-3	$\sigma$&&	1	\FiveStarOpen&&	2	\FiveStarOpen\\[.02cm]
 	\cirn{2}~\cirn{3}~\cirn{8}&&	0.405&&	0.300&&	0.496&&	5.4863 	&&	6.44$\TIMES10^{-2}$&&	1-2	$\sigma$&&	4	\FiveStarOpen&&	4	\FiveStarOpen\\[.02cm]
 	\cirn{2}~\cirn{3}~\cirn{9}&&	0.733&&	0.260&&	0.412&&	5.0196 	&&	8.13$\TIMES10^{-2}$&&	1-2	$\sigma$&&	4	\FiveStarOpen&&	3	\FiveStarOpen\\[.02cm]
 	\cirn{2}~\cirn{4}~\cirn{5}&&	-0.014&&0.320&&	0.339&&	2.2364 	&&	3.27$\TIMES10^{-1}$&&	$<$1$\sigma$&&	5	\FiveStarOpen&&	4	\FiveStarOpen\\[.02cm]
 	\cirn{2}~\cirn{4}~\cirn{7}&&	-1.145&&0.120&&	0.298&&	12.2685 &&	2.17$\TIMES10^{-3}$&&	3-4	$\sigma$&&	1	\FiveStarOpen&&	2	\FiveStarOpen\\[.02cm]
 	\cirn{2}~\cirn{4}~\cirn{8}&&	-0.168&&0.304&&	0.349&&	2.6384 	&&	2.67$\TIMES10^{-1}$&&	1-2	$\sigma$&&	4	\FiveStarOpen&&	4	\FiveStarOpen\\[.02cm]
 	\cirn{2}~\cirn{4}~\cirn{9}&&	0.165&&	0.291&&	0.554&&	7.2437 	&&	2.67$\TIMES10^{-2}$&&	2-3	$\sigma$&&	4	\FiveStarOpen&&	4	\FiveStarOpen\\[.02cm]
 	\cirn{2}~\cirn{4}~\cirn{12}&&	-1.517&&0.219&&	0.783&&	25.6687 &&	2.67$\TIMES10^{-6}$&&	4-5	$\sigma$&&	1	\FiveStarOpen&&	2	\FiveStarOpen\\[.02cm]
 	\cirn{2}~\cirn{4}~\cirn{15}&&	-1.517&&0.219&&	0.783&&	25.6687 &&	2.67$\TIMES10^{-6}$&&	4-5	$\sigma$&&	1	\FiveStarOpen&&	2	\FiveStarOpen\\[.02cm]
 	\cirn{2}~\cirn{5}~\cirn{6}&&	1.178&&	0.169&&	0.377&&	9.9818 	&&	6.80$\TIMES10^{-3}$&&	2-3	$\sigma$&&	1	\FiveStarOpen&&	2	\FiveStarOpen\\[.02cm]
 	\cirn{2}~\cirn{5}~\cirn{7}&&	-1.176&&0.127&&	0.297&&	10.9280	&&	4.24$\TIMES10^{-3}$&&	2-3	$\sigma$&&	1	\FiveStarOpen&&	2	\FiveStarOpen\\[.02cm]
 	\cirn{2}~\cirn{5}~\cirn{8}&&	-0.023&&0.417&&	0.534&&	3.2771 	&&	1.94$\TIMES10^{-1}$&&	1-2	$\sigma$&&	4	\FiveStarOpen&&	4	\FiveStarOpen\\[.02cm]
 	\cirn{2}~\cirn{5}~\cirn{9}&&	0.635&&	0.316&&	0.490&&	4.8104 	&&	9.02$\TIMES10^{-2}$&&	1-2	$\sigma$&&	4	\FiveStarOpen&&	3	\FiveStarOpen\\[.02cm]
 	\cirn{2}~\cirn{6}~\cirn{9}&&	1.210&&	0.168&&	0.401&&	11.4751 &&	3.22$\TIMES10^{-3}$&&	2-3	$\sigma$&&	1	\FiveStarOpen&&	2	\FiveStarOpen\\[.02cm]
 	\cirn{2}~\cirn{7}~\cirn{8}&&	-1.172&&0.125&&	0.283&&	10.2145 &&	6.05$\TIMES10^{-3}$&&	2-3	$\sigma$&&	1	\FiveStarOpen&&	2	\FiveStarOpen\\[.02cm]
 	\cirn{2}~\cirn{7}~\cirn{9}&&	-1.144&&0.122&&	0.397&&	21.0808	&&	2.64$\TIMES10^{-5}$&&	4-5	$\sigma$&&	1	\FiveStarOpen&&	2	\FiveStarOpen\\[.02cm]
 	\cirn{2}~\cirn{7}~\cirn{12}&&	-1.423&&0.137&&	0.465&&	23.0357	&&	9.95$\TIMES10^{-6}$&&	4-5	$\sigma$&&	0	\FiveStarOpen&&	2	\FiveStarOpen\\[.02cm]
 	\cirn{2}~\cirn{7}~\cirn{15}&&	-1.423&&0.137&&	0.465&&	23.0357 &&	9.95$\TIMES10^{-6}$&&	4-5	$\sigma$&&	0	\FiveStarOpen&&	2	\FiveStarOpen\\[.02cm]
 	\cirn{2}~\cirn{8}~\cirn{9}&&	0.320&&	0.346&&	0.690&&	7.9486 	&&	1.88$\TIMES10^{-2}$&&	2-3	$\sigma$&&	4	\FiveStarOpen&&	3	\FiveStarOpen\\[.02cm]
 	\cirn{2}~\cirn{12}~\cirn{15}&&-1.947&& 0.175&& 0.594&& 22.9756&&	1.03$\TIMES10^{-5}$&&	4-5	$\sigma$&&	0	\FiveStarOpen&&	1	\FiveStarOpen\\[.02cm]
 	\cirn{3}~\cirn{4}~\cirn{5}&&	0.134&&	0.328&&	0.544&&	5.5008 	&&	6.39$\TIMES10^{-2}$&&	1-2	$\sigma$&&	4	\FiveStarOpen&&	4	\FiveStarOpen\\[.02cm]
 	\cirn{3}~\cirn{4}~\cirn{8}&&	-0.037&&0.306&&	0.548&&	6.4105 	&&	4.05$\TIMES10^{-2}$&&	2-3	$\sigma$&&	5	\FiveStarOpen&&	4	\FiveStarOpen\\[.02cm]
 	\cirn{3}~\cirn{4}~\cirn{9}&&	0.312&&	0.319&&	0.704&&	9.7672 	&&	7.57$\TIMES10^{-3}$&&	2-3	$\sigma$&&	4	\FiveStarOpen&&	3	\FiveStarOpen\\[.02cm]
 	\cirn{3}~\cirn{5}~\cirn{6}&&	1.511&&	0.168&&	0.168&&	1.1644 	&&	5.59$\TIMES10^{-1}$&&	$<$1$\sigma$&&	0	\FiveStarOpen&&	0	\FiveStarOpen\\[.02cm]
 	\cirn{3}~\cirn{5}~\cirn{8}&&	0.306&&	0.459&&	0.810&&	6.2152 	&&	4.47$\TIMES10^{-2}$&&	2-3	$\sigma$&&	4	\FiveStarOpen&&	3	\FiveStarOpen\\[.02cm]
 	\cirn{3}~\cirn{5}~\cirn{9}&&	1.469&&	0.425&&	0.496&&	2.7226 	&&	2.56$\TIMES10^{-1}$&&	1-2	$\sigma$&&	1	\FiveStarOpen&&	2	\FiveStarOpen\\[.02cm]
 	\cirn{3}~\cirn{5}~\cirn{13}&&	1.768&&	0.141&&	0.158&&	2.5237 	&&	2.83$\TIMES10^{-1}$&&	1-2	$\sigma$&&	0	\FiveStarOpen&&	0	\FiveStarOpen\\[.02cm]
 	\cirn{3}~\cirn{5}~\cirn{14}&&	1.768&&	0.141&&	0.158&&	2.5237 	&&	2.83$\TIMES10^{-1}$&&	1-2	$\sigma$&&	0	\FiveStarOpen&&	0	\FiveStarOpen\\[.02cm]
 	\cirn{3}~\cirn{6}~\cirn{8}&&	1.334&&	0.185&&	0.443&&	11.5077 &&	3.17$\TIMES10^{-3}$&&	2-3	$\sigma$&&	1	\FiveStarOpen&&	2	\FiveStarOpen\\[.02cm]
 	\cirn{3}~\cirn{6}~\cirn{9}&&	1.538&&	0.163&&	0.163&&	1.9694	&&	3.74$\TIMES10^{-1}$&&	$<$1$\sigma$&&	0	\FiveStarOpen&&	0	\FiveStarOpen\\[.02cm]
 	\cirn{3}~\cirn{6}~\cirn{13}&&	1.687&&	0.098&&	0.121&&	3.0709 	&&	2.15$\TIMES10^{-1}$&&	1-2	$\sigma$&&	0	\FiveStarOpen&&	0	\FiveStarOpen\\[.02cm]
 	\cirn{3}~\cirn{6}~\cirn{14}&&	1.687&&	0.098&&	0.121&&	3.0709 	&&	2.15$\TIMES10^{-1}$&&	1-2	$\sigma$&&	0	\FiveStarOpen&&	0	\FiveStarOpen\\[.02cm]
 	\cirn{3}~\cirn{8}~\cirn{9}&&	0.619&&	0.434&&	0.955&&	9.6864 	&&	7.88$\TIMES10^{-3}$&&	2-3	$\sigma$&&	3	\FiveStarOpen&&	3	\FiveStarOpen\\[.02cm]
 	\cirn{3}~\cirn{9}~\cirn{13}&&	1.782&&	0.143&&	0.171&&	2.8672 	&&	2.38$\TIMES10^{-1}$&&	1-2	$\sigma$&&	0	\FiveStarOpen&&	0	\FiveStarOpen\\[.02cm]
 	\cirn{3}~\cirn{9}~\cirn{14}&&	1.782&&	0.143&&	0.171&&	2.8672 	&&	2.38$\TIMES10^{-1}$&&	1-2	$\sigma$&&	0	\FiveStarOpen&&	0	\FiveStarOpen\\[.02cm]
 	\cirn{3}~\cirn{13}~\cirn{14}&& 1.796&& 0.107&&0.107&&	1.8123 	&&	4.04$\TIMES10^{-1}$&&	$<$1$\sigma$&&	0	\FiveStarOpen&&	0	\FiveStarOpen\\[.02cm]
 	\cirn{4}~\cirn{5}~\cirn{7}&&	-1.224&&0.134&&	0.174&&	3.3680 	&&	1.86$\TIMES10^{-1}$&&	1-2	$\sigma$&&	1	\FiveStarOpen&&	1	\FiveStarOpen\\[.02cm]
 	\cirn{4}~\cirn{5}~\cirn{8}&&	-0.479&&0.420&&	0.420&&	1.1892 	&&	5.52$\TIMES10^{-1}$&&	$<$1$\sigma$&&	4	\FiveStarOpen&&	4	\FiveStarOpen\\[.02cm]
 	\cirn{4}~\cirn{5}~\cirn{9}&&	-0.009&&0.387&&	0.738&&	7.2735 	&&	2.63$\TIMES10^{-2}$&&	2-3	$\sigma$&&	4	\FiveStarOpen&&	4	\FiveStarOpen\\[.02cm]
 	\cirn{4}~\cirn{5}~\cirn{12}&&	-1.916&&0.239&&	0.530&&	9.8071 	&&	7.42$\TIMES10^{-3}$&&	2-3	$\sigma$&&	0	\FiveStarOpen&&	1	\FiveStarOpen\\[.02cm]
 	\cirn{4}~\cirn{5}~\cirn{15}&&	-1.916&&0.239&&	0.530&&	9.8071 	&&	7.42$\TIMES10^{-3}$&&	2-3	$\sigma$&&	0	\FiveStarOpen&&	1	\FiveStarOpen\\[.02cm]
 	\cirn{4}~\cirn{7}~\cirn{8}&&	-1.218&&0.132&&	0.153&&	2.6786 	&&	2.62$\TIMES10^{-1}$&&	1-2	$\sigma$&&	1	\FiveStarOpen&&	1	\FiveStarOpen\\[.02cm]
 	\cirn{4}~\cirn{7}~\cirn{9}&&	-1.186&&0.128&&	0.336&&	13.7154 &&	1.05$\TIMES10^{-3}$&&	3-4	$\sigma$&&	1	\FiveStarOpen&&	2	\FiveStarOpen\\[.02cm]
 	\cirn{4}~\cirn{7}~\cirn{12}&&	-1.493&&0.140&&	0.366&&	13.7339 &&	1.04$\TIMES10^{-3}$&&	3-4	$\sigma$&&	0	\FiveStarOpen&&	1	\FiveStarOpen\\[.02cm]
 	\cirn{4}~\cirn{7}~\cirn{15}&&	-1.493&&0.140&&	0.366&&	13.7339 &&	1.04$\TIMES10^{-3}$&&	3-4	$\sigma$&&	0	\FiveStarOpen&&	1	\FiveStarOpen\\[.02cm]
 	\cirn{4}~\cirn{8}~\cirn{9}&&	-0.205&&0.360&&	0.704&&	7.6597 	&&	2.17$\TIMES10^{-2}$&&	2-3	$\sigma$&&	4	\FiveStarOpen&&	4	\FiveStarOpen\\[.02cm]
 	\cirn{4}~\cirn{8}~\cirn{12}&&	-1.876&&0.235&&	0.518&&	9.7241 	&&	7.73$\TIMES10^{-3}$&&	2-3	$\sigma$&&	0	\FiveStarOpen&&	1	\FiveStarOpen\\[.02cm]
 	\cirn{4}~\cirn{8}~\cirn{15}&&	-1.876&&0.235&&	0.518&&	9.7241 	&&	7.73$\TIMES10^{-3}$&&	2-3	$\sigma$&&	0	\FiveStarOpen&&	1	\FiveStarOpen\\[.02cm]
 	\cirn{4}~\cirn{9}~\cirn{12}&&	-1.753&&0.236&&	0.808&&	23.5452 &&	7.71$\TIMES10^{-6}$&&	4-5	$\sigma$&&	0	\FiveStarOpen&&	2	\FiveStarOpen\\[.02cm]
 	\cirn{4}~\cirn{9}~\cirn{15}&&	-1.753&&0.236&&	0.808&&	23.5452 &&	7.71$\TIMES10^{-6}$&&	4-5	$\sigma$&&	0	\FiveStarOpen&&	2	\FiveStarOpen\\[.02cm]
 	\cirn{4}~\cirn{12}~\cirn{15}&&-2.095&&0.160&&	0.357&&	9.9977 	&&	6.75$\TIMES10^{-3}$&&	2-3	$\sigma$&&	0	\FiveStarOpen&&	0	\FiveStarOpen\\[.02cm]
 	\cirn{5}~\cirn{6}~\cirn{9}&&	1.579&&	0.169&&	0.182&&	2.3369 	&&	3.11$\TIMES10^{-1}$&&	1-2	$\sigma$&&	0	\FiveStarOpen&&	0	\FiveStarOpen\\[.02cm]
 	\cirn{5}~\cirn{6}~\cirn{13}&&	1.708&&	0.101&&	0.118&&	2.7765 	&&	2.50$\TIMES10^{-1}$&&	1-2	$\sigma$&&	0	\FiveStarOpen&&	0	\FiveStarOpen\\[.02cm]
 	\cirn{5}~\cirn{6}~\cirn{14}&&	1.708&&	0.101&&	0.118&&	2.7765 	&&	2.50$\TIMES10^{-1}$&&	1-2	$\sigma$&&	0	\FiveStarOpen&&	0	\FiveStarOpen\\[.02cm]
 	\cirn{5}~\cirn{7}~\cirn{8}&&	-1.267&&0.143&&	0.143&&	0.9609 	&&	6.18$\TIMES10^{-1}$&&	$<$1$\sigma$&&	1	\FiveStarOpen&&	1	\FiveStarOpen\\[.02cm]
 	\cirn{5}~\cirn{7}~\cirn{9}&&	-1.226&&0.137&&	0.339&&	12.1760 &&	2.27$\TIMES10^{-3}$&&	3-4	$\sigma$&&	1	\FiveStarOpen&&	2	\FiveStarOpen\\[.02cm]
 	\cirn{5}~\cirn{7}~\cirn{12}&&	-1.565&&0.144&&	0.326&&	10.2336 &&	6.00$\TIMES10^{-3}$&&	2-3	$\sigma$&&	0	\FiveStarOpen&&	1	\FiveStarOpen\\[.02cm]
 	\cirn{5}~\cirn{7}~\cirn{15}&&	-1.565&&0.144&&	0.326&&	10.2336 &&	6.00$\TIMES10^{-3}$&&	2-3	$\sigma$&&	0	\FiveStarOpen&&	1	\FiveStarOpen\\[.02cm]
 	\cirn{5}~\cirn{8}~\cirn{9}&&	-0.011&&0.609&&	1.241&&	8.3210 	&&	1.56$\TIMES10^{-2}$&&	2-3	$\sigma$&&	4	\FiveStarOpen&&	3	\FiveStarOpen\\[.02cm]
 	\cirn{5}~\cirn{8}~\cirn{12}&&	-2.202&&0.207&&	0.207&&	1.9114 	&&	3.85$\TIMES10^{-1}$&&	$<$1$\sigma$&&	0	\FiveStarOpen&&	0	\FiveStarOpen\\[.02cm]
 	\cirn{5}~\cirn{8}~\cirn{15}&&	-2.202&&0.207&&	0.207&&	1.9114 	&&	3.85$\TIMES10^{-1}$&&	$<$1$\sigma$&&	0	\FiveStarOpen&&	0	\FiveStarOpen\\[.02cm]
 	\cirn{5}~\cirn{9}~\cirn{12}&&	-2.060&&0.239&&	0.701&&	17.2299 &&	1.81$\TIMES10^{-4}$&&	3-4	$\sigma$&&	0	\FiveStarOpen&&	1	\FiveStarOpen\\[.02cm]
 	\cirn{5}~\cirn{9}~\cirn{13}&&	1.859&&	0.171&&	0.171&&	1.8552 	&&	3.96$\TIMES10^{-1}$&&	$<$1$\sigma$&&	0	\FiveStarOpen&&	0	\FiveStarOpen\\[.02cm]
 	\cirn{5}~\cirn{9}~\cirn{14}&&	1.859&&	0.171&&	0.171&&	1.8552 	&&	3.96$\TIMES10^{-1}$&&	$<$1$\sigma$&&	0	\FiveStarOpen&&	0	\FiveStarOpen\\[.02cm]
 	\cirn{5}~\cirn{9}~\cirn{15}&&	-2.060&&0.239&&	0.701&&	17.2299 &&	1.81$\TIMES10^{-4}$&&	3-4	$\sigma$&&	0	\FiveStarOpen&&	1	\FiveStarOpen\\[.02cm]
 	\cirn{5}~\cirn{12}~\cirn{15}&&-2.238&&0.142&&	0.142&&	0.8100 	&&	6.67$\TIMES10^{-1}$&&	$<$1$\sigma$&&	0	\FiveStarOpen&&	0	\FiveStarOpen\\[.02cm]
 	\cirn{5}~\cirn{13}~\cirn{14}&&1.836&&	0.118&&	0.118&&	0.8100 	&&	6.67$\TIMES10^{-1}$&&	$<$1$\sigma$&&	0	\FiveStarOpen&&	0	\FiveStarOpen\\[.02cm]
 	\cirn{6}~\cirn{9}~\cirn{13}&&	1.715&&	0.101&&	0.128&&	3.2120 	&&	2.01$\TIMES10^{-1}$&&	1-2	$\sigma$&&	0	\FiveStarOpen&&	0	\FiveStarOpen\\[.02cm]
 	\cirn{6}~\cirn{9}~\cirn{14}&&	1.715&&	0.101&&	0.128&&	3.2120 	&&	2.01$\TIMES10^{-1}$&&	1-2	$\sigma$&&	0	\FiveStarOpen&&	0	\FiveStarOpen\\[.02cm]
 	\cirn{6}~\cirn{13}~\cirn{14}&& 1.746&&0.085&&	0.093&&	2.4398 	&&	2.95$\TIMES10^{-1}$&&	1-2	$\sigma$&&	0	\FiveStarOpen&&	0	\FiveStarOpen\\[.02cm]
 	\cirn{7}~\cirn{8}~\cirn{9}&&	-1.220&&0.135&&	0.324&&	11.4877 &&	3.20$\TIMES10^{-3}$&&	2-3	$\sigma$&&	1	\FiveStarOpen&&	2	\FiveStarOpen\\[.02cm]
 	\cirn{7}~\cirn{8}~\cirn{12}&&	-1.553&&0.143&&	0.316&&	9.8089 	&&	7.41$\TIMES10^{-3}$&&	2-3	$\sigma$&&	0	\FiveStarOpen&&	1	\FiveStarOpen\\[.02cm]
 	\cirn{7}~\cirn{8}~\cirn{15}&&	-1.553&&0.143&&	0.316&&	9.8089 	&&	7.41$\TIMES10^{-3}$&&	2-3	$\sigma$&&	0	\FiveStarOpen&&	1	\FiveStarOpen\\[.02cm]
 	\cirn{7}~\cirn{9}~\cirn{12}&&	-1.509&&0.143&&	0.478&&	22.3006 &&	1.44$\TIMES10^{-5}$&&	4-5	$\sigma$&&	0	\FiveStarOpen&&	1	\FiveStarOpen\\[.02cm]
 	\cirn{7}~\cirn{9}~\cirn{15}&&	-1.509&&0.143&&	0.478&&	22.3006 &&	1.44$\TIMES10^{-5}$&&	4-5	$\sigma$&&	0	\FiveStarOpen&&	1	\FiveStarOpen\\[.02cm]
 	\cirn{7}~\cirn{12}~\cirn{15}&&-1.720&&0.126&&	0.340&&	14.4854 &&	7.15$\TIMES10^{-4}$&&	3-4	$\sigma$&&	0	\FiveStarOpen&&	1	\FiveStarOpen\\[.02cm]
 	\cirn{8}~\cirn{12}~\cirn{15}&&-2.220&&0.144&&	0.144&&	1.1166 	&&	5.72$\TIMES10^{-1}$&&	$<$1$\sigma$&&	0	\FiveStarOpen&&	0	\FiveStarOpen\\[.02cm]
 	\cirn{9}~\cirn{12}~\cirn{15}&&-2.164&&0.153&&	0.444&&	16.7466 &&	2.31$\TIMES10^{-4}$&&	3-4	$\sigma$&&	0	\FiveStarOpen&&	1	\FiveStarOpen\\[.02cm]
 	\cirn{9}~\cirn{13}~\cirn{14}&&1.847&&	0.119&&	0.119&&	1.0543 	&&	5.90$\TIMES10^{-1}$&&	$<$1$\sigma$&&	0	\FiveStarOpen&&	0	\FiveStarOpen\\[.02cm]
\end{longtable*}

\begin{longtable*}{ccccccccccccccccccccccccc}
\caption{\label{tab:tripair_IH}
Tripair combinations in IH.}\\
\hline\\[-3mm]
\hline\\[-.3cm]
\multirow{2}{*}{Pairs included}&&\multirow{2}{*}{$\overline{\cos\delta}$}&&\multicolumn{3}{c}{$\sigma_{\overline{\cos\delta}}$}&&\multicolumn{5}{c}{Self consistency}&&\multicolumn{3}{c}{Natural limit consistency}\\
\cline{5-7}\cline{9-13}\cline{15-17}\\[-.27cm]
&&&&Unscaled&&Scaled&&$\chi^2$&&p-value&&Exclusion level&&Unscaled&&Scaled\\
\\[-.4cm] \hline
\endhead
\hline \multicolumn{18  }{r}{{Continued on next page}} \\ \hline \hline
\endfoot
\hline \hline
\endlastfoot
 	\cirn{1}~\cirn{2}~\cirn{3}&&	0.146&&	0.145&&	0.273   &&	7.0365 	&&	2.97$\TIMES10^{-2}$&&	2-3	$\sigma$&&	5 \FiveStarOpen&&	5 \FiveStarOpen\\[.02cm]
 	\cirn{1}~\cirn{2}~\cirn{4}&&	0.018&&	0.148&&	0.441 	&&	17.7885 &&	1.37$\TIMES10^{-4}$&&	3-4	$\sigma$&&	5 \FiveStarOpen&&	4 \FiveStarOpen\\[.02cm]
 	\cirn{1}~\cirn{2}~\cirn{5}&&	0.166&&	0.155&&	0.364 	&&	11.0783 &&	3.93$\TIMES10^{-3}$&&	2-3	$\sigma$&&	5 \FiveStarOpen&&	4 \FiveStarOpen\\[.02cm]
 	\cirn{1}~\cirn{2}~\cirn{8}&&	0.068&&	0.153&&	0.528 	&&	23.9351 &&	6.35$\TIMES10^{-6}$&&	4-5	$\sigma$&&	5 \FiveStarOpen&&	4 \FiveStarOpen\\[.02cm]
 	\cirn{1}~\cirn{2}~\cirn{9}&&	0.164&&	0.152&&	0.282 	&&	6.8992 	&&	3.18$\TIMES10^{-2}$&&	2-3	$\sigma$&&	5 \FiveStarOpen&&	4 \FiveStarOpen\\[.02cm]
 	\cirn{1}~\cirn{3}~\cirn{4}&&	0.130&&	0.149&&	0.381 	&&	13.0255 &&	1.48$\TIMES10^{-3}$&&	3-4	$\sigma$&&	5 \FiveStarOpen&&	4 \FiveStarOpen\\[.02cm]
 	\cirn{1}~\cirn{3}~\cirn{5}&&	0.305&&	0.135&&	0.208 	&&	4.7194 	&&	9.44$\TIMES10^{-2}$&&	1-2	$\sigma$&&	5 \FiveStarOpen&&	5 \FiveStarOpen\\[.02cm]
 	\cirn{1}~\cirn{3}~\cirn{8}&&	0.196&&	0.150&&	0.460 	&&	18.6676 &&	8.84$\TIMES10^{-5}$&&	3-4	$\sigma$&&	5 \FiveStarOpen&&	4 \FiveStarOpen\\[.02cm]
 	\cirn{1}~\cirn{3}~\cirn{9}&&	0.299&&	0.134&&	0.134 	&&	0.5921 	&&	7.44$\TIMES10^{-1}$&&	$<$1$\sigma$&&	5 \FiveStarOpen&&	5 \FiveStarOpen\\[.02cm]
 	\cirn{1}~\cirn{4}~\cirn{5}&&	0.149&&	0.159&&	0.466 	&&	17.0946 &&	1.94$\TIMES10^{-4}$&&	3-4	$\sigma$&&	5 \FiveStarOpen&&	4 \FiveStarOpen\\[.02cm]
 	\cirn{1}~\cirn{4}~\cirn{8}&&	0.045&&	0.157&&	0.607 	&&	29.7848 &&	3.41$\TIMES10^{-7}$&&	$>$5$\sigma$&&	5 \FiveStarOpen&&	4 \FiveStarOpen\\[.02cm]
 	\cirn{1}~\cirn{4}~\cirn{9}&&	0.147&&	0.156&&	0.397 	&&	12.9129 &&	1.57$\TIMES10^{-3}$&&	3-4	$\sigma$&&	5 \FiveStarOpen&&	4 \FiveStarOpen\\[.02cm]
 	\cirn{1}~\cirn{5}~\cirn{8}&&	0.233&&	0.154&&	0.517 	&&	22.6448 &&	1.21$\TIMES10^{-5}$&&	4-5	$\sigma$&&	5 \FiveStarOpen&&	4 \FiveStarOpen\\[.02cm]
 	\cirn{1}~\cirn{5}~\cirn{9}&&	0.330&&	0.136&&	0.198 	&&	4.2590 	&&	1.19$\TIMES10^{-1}$&&	1-2	$\sigma$&&	5 \FiveStarOpen&&	5 \FiveStarOpen\\[.02cm]
 	\cirn{1}~\cirn{8}~\cirn{9}&&	0.227&&	0.152&&	0.462 	&&	18.4770 &&	9.72$\TIMES10^{-5}$&&	3-4	$\sigma$&&	5 \FiveStarOpen&&	4 \FiveStarOpen\\[.02cm]
 	\cirn{2}~\cirn{3}~\cirn{4}&&	-0.657&&0.228&&	0.416 	&&	6.6486 	&&	3.60$\TIMES10^{-2}$&&	2-3	$\sigma$&&	4 \FiveStarOpen&&	3 \FiveStarOpen\\[.02cm]
 	\cirn{2}~\cirn{3}~\cirn{5}&&	-0.373&&0.272&&	0.483 	&&	6.3183 	&&	4.25$\TIMES10^{-2}$&&	2-3	$\sigma$&&	4 \FiveStarOpen&&	4 \FiveStarOpen\\[.02cm]
 	\cirn{2}~\cirn{3}~\cirn{6}&&	0.978&&	0.108&&	0.403 	&&	27.7377 &&	9.48$\TIMES10^{-7}$&&	4-5	$\sigma$&&	3 \FiveStarOpen&&	3 \FiveStarOpen\\[.02cm]
 	\cirn{2}~\cirn{3}~\cirn{8}&&	-0.642&&0.248&&	0.667 	&&	14.4852 &&	7.15$\TIMES10^{-4}$&&	3-4	$\sigma$&&	4 \FiveStarOpen&&	3 \FiveStarOpen\\[.02cm]
 	\cirn{2}~\cirn{3}~\cirn{9}&&	-0.313&&0.255&&	0.287 	&&	2.5383 	&&	2.81$\TIMES10^{-1}$&&	1-2	$\sigma$&&	4 \FiveStarOpen&&	4 \FiveStarOpen\\[.02cm]
 	\cirn{2}~\cirn{4}~\cirn{5}&&	-0.994&&0.251&&	0.436 	&&	6.0217 	&&	4.92$\TIMES10^{-2}$&&	1-2	$\sigma$&&	3 \FiveStarOpen&&	3 \FiveStarOpen\\[.02cm]
 	\cirn{2}~\cirn{4}~\cirn{7}&&	-1.404&&0.147&&	0.271 	&&	6.8407 	&&	3.27$\TIMES10^{-2}$&&	2-3	$\sigma$&&	1 \FiveStarOpen&&	1 \FiveStarOpen\\[.02cm]
 	\cirn{2}~\cirn{4}~\cirn{8}&&	-1.148&&0.234&&	0.537 	&&	10.5627 &&	5.09$\TIMES10^{-3}$&&	2-3	$\sigma$&&	2 \FiveStarOpen&&	2 \FiveStarOpen\\[.02cm]
 	\cirn{2}~\cirn{4}~\cirn{9}&&	-0.855&&0.248&&	0.366 	&&	4.3594 	&&	1.13$\TIMES10^{-1}$&&	1-2	$\sigma$&&	3 \FiveStarOpen&&	3 \FiveStarOpen\\[.02cm]
 	\cirn{2}~\cirn{4}~\cirn{12}&&	-1.565&&0.108&&	0.233 	&&	9.3169 	&&	9.48$\TIMES10^{-3}$&&	2-3	$\sigma$&&	0 \FiveStarOpen&&	1 \FiveStarOpen\\[.02cm]
 	\cirn{2}~\cirn{4}~\cirn{15}&&	-1.565&&0.108&&	0.233 	&&	9.3169 	&&	9.48$\TIMES10^{-3}$&&	2-3	$\sigma$&&	0 \FiveStarOpen&&	1 \FiveStarOpen\\[.02cm]
 	\cirn{2}~\cirn{5}~\cirn{6}&&	1.036&&	0.111&&	0.410 	&&	27.1098 &&	1.30$\TIMES10^{-6}$&&	4-5	$\sigma$&&	2 \FiveStarOpen&&	2 \FiveStarOpen\\[.02cm]
 	\cirn{2}~\cirn{5}~\cirn{7}&&	-1.383&&0.154&&	0.359 	&&	10.8413 &&	4.42$\TIMES10^{-3}$&&	2-3	$\sigma$&&	1 \FiveStarOpen&&	1 \FiveStarOpen\\[.02cm]
 	\cirn{2}~\cirn{5}~\cirn{8}&&	-1.036&&0.270&&	0.707 	&&	13.6998 &&	1.06$\TIMES10^{-3}$&&	3-4	$\sigma$&&	2 \FiveStarOpen&&	2 \FiveStarOpen\\[.02cm]
 	\cirn{2}~\cirn{5}~\cirn{9}&&	-0.581&&0.319&&	0.537 	&&	5.6874 	&&	5.82$\TIMES10^{-2}$&&	1-2	$\sigma$&&	4 \FiveStarOpen&&	3 \FiveStarOpen\\[.02cm]
 	\cirn{2}~\cirn{6}~\cirn{9}&&	1.023&&	0.111&&	0.382 	&&	23.8825 &&	6.52$\TIMES10^{-6}$&&	4-5	$\sigma$&&	2 \FiveStarOpen&&	2 \FiveStarOpen\\[.02cm]
 	\cirn{2}~\cirn{7}~\cirn{8}&&	-1.445&&0.152&&	0.394 	&&	13.3598 &&	1.26$\TIMES10^{-3}$&&	3-4	$\sigma$&&	1 \FiveStarOpen&&	1 \FiveStarOpen\\[.02cm]
 	\cirn{2}~\cirn{7}~\cirn{9}&&	-1.303&&0.150&&	0.363 	&&	11.6918 &&	2.89$\TIMES10^{-3}$&&	2-3	$\sigma$&&	1 \FiveStarOpen&&	2 \FiveStarOpen\\[.02cm]
 	\cirn{2}~\cirn{7}~\cirn{12}&&	-1.590&&0.094&&	0.205 	&&	9.5075 	&&	8.62$\TIMES10^{-3}$&&	2-3	$\sigma$&&	0 \FiveStarOpen&&	1 \FiveStarOpen\\[.02cm]
 	\cirn{2}~\cirn{7}~\cirn{15}&&	-1.590&&0.094&&	0.205 	&&	9.5075 	&&	8.62$\TIMES10^{-3}$&&	2-3	$\sigma$&&	0 \FiveStarOpen&&	1 \FiveStarOpen\\[.02cm]
 	\cirn{2}~\cirn{8}~\cirn{9}&&	-0.878&&0.269&&	0.665 	&&	12.1759 &&	2.27$\TIMES10^{-3}$&&	3-4	$\sigma$&&	3 \FiveStarOpen&&	3 \FiveStarOpen\\[.02cm]
 	\cirn{2}~\cirn{12}~\cirn{15}&&-1.626&&0.085&&	0.190 	&&	10.0760 &&	6.49$\TIMES10^{-3}$&&	2-3	$\sigma$&&	0 \FiveStarOpen&&	0 \FiveStarOpen\\[.02cm]
 	\cirn{3}~\cirn{4}~\cirn{5}&&	-0.553&&0.300&&	0.688 	&&	10.5431 &&	5.14$\TIMES10^{-3}$&&	2-3	$\sigma$&&	4 \FiveStarOpen&&	3 \FiveStarOpen\\[.02cm]
 	\cirn{3}~\cirn{4}~\cirn{8}&&	-0.839&&0.287&&	0.844 	&&	17.3020 &&	1.75$\TIMES10^{-4}$&&	3-4	$\sigma$&&	3 \FiveStarOpen&&	3 \FiveStarOpen\\[.02cm]
 	\cirn{3}~\cirn{4}~\cirn{9}&&	-0.441&&0.273&&	0.518 	&&	7.1859 	&&	2.75$\TIMES10^{-2}$&&	2-3	$\sigma$&&	4 \FiveStarOpen&&	4 \FiveStarOpen\\[.02cm]
 	\cirn{3}~\cirn{5}~\cirn{6}&&	1.116&&	0.115&&	0.264 	&&	10.4852 &&	5.29$\TIMES10^{-3}$&&	2-3	$\sigma$&&	1 \FiveStarOpen&&	2 \FiveStarOpen\\[.02cm]
 	\cirn{3}~\cirn{5}~\cirn{8}&&	-0.490&&0.347&&	1.049 	&&	18.2751 &&	1.08$\TIMES10^{-4}$&&	3-4	$\sigma$&&	4 \FiveStarOpen&&	3 \FiveStarOpen\\[.02cm]
 	\cirn{3}~\cirn{5}~\cirn{9}&&	0.022&&	0.362&&	0.525 	&&	4.2024 	&&	1.22$\TIMES10^{-1}$&&	1-2	$\sigma$&&	4 \FiveStarOpen&&	4 \FiveStarOpen\\[.02cm]
 	\cirn{3}~\cirn{5}~\cirn{13}&&	1.954&&	0.196&&	0.710 	&&	26.2075 &&	2.04$\TIMES10^{-6}$&&	4-5	$\sigma$&&	0 \FiveStarOpen&&	1 \FiveStarOpen\\[.02cm]
 	\cirn{3}~\cirn{5}~\cirn{14}&&	1.954&&	0.196&&	0.710 	&&	26.2075 &&	2.04$\TIMES10^{-6}$&&	4-5	$\sigma$&&	0 \FiveStarOpen&&	1 \FiveStarOpen\\[.02cm]
 	\cirn{3}~\cirn{6}~\cirn{8}&&	1.055&&	0.112&&	0.450 	&&	32.3693 &&	9.36$\TIMES10^{-8}$&&	$>$5$\sigma$&&	2 \FiveStarOpen&&	2 \FiveStarOpen\\[.02cm]
 	\cirn{3}~\cirn{6}~\cirn{9}&&	1.102&&	0.114&&	0.219 	&&	7.4249 	&&	2.44$\TIMES10^{-2}$&&	2-3	$\sigma$&&	2 \FiveStarOpen&&	2 \FiveStarOpen\\[.02cm]
 	\cirn{3}~\cirn{6}~\cirn{13}&&	1.435&&	0.111&&	0.446 	&&	32.3245 &&	9.57$\TIMES10^{-8}$&&	$>$5$\sigma$&&	0 \FiveStarOpen&&	2 \FiveStarOpen\\[.02cm]
 	\cirn{3}~\cirn{6}~\cirn{14}&&	1.435&&	0.111&&	0.446 	&&	32.3245 &&	9.57$\TIMES10^{-8}$&&	$>$5$\sigma$&&	0 \FiveStarOpen&&	2 \FiveStarOpen\\[.02cm]
 	\cirn{3}~\cirn{8}~\cirn{9}&&	-0.367&&0.310&&	0.839 	&&	14.7075 &&	6.40$\TIMES10^{-4}$&&	3-4	$\sigma$&&	4 \FiveStarOpen&&	3 \FiveStarOpen\\[.02cm]
 	\cirn{3}~\cirn{9}~\cirn{13}&&	1.878&&	0.192&&	0.688 	&&	25.6714 &&	2.66$\TIMES10^{-6}$&&	4-5	$\sigma$&&	0 \FiveStarOpen&&	1 \FiveStarOpen\\[.02cm]
 	\cirn{3}~\cirn{9}~\cirn{14}&&	1.878&&	0.192&&	0.688 	&&	25.6714 &&	2.66$\TIMES10^{-6}$&&	4-5	$\sigma$&&	0 \FiveStarOpen&&	1 \FiveStarOpen\\[.02cm]
 	\cirn{3}~\cirn{13}~\cirn{14}&&2.149&&	0.145&&	0.506 	&&	24.2126 &&	5.52$\TIMES10^{-6}$&&	4-5	$\sigma$&&	0 \FiveStarOpen&&	1 \FiveStarOpen\\[.02cm]
 	\cirn{4}~\cirn{5}~\cirn{7}&&	-1.648&&0.160&&	0.231 	&&	4.1987 	&&	1.23$\TIMES10^{-1}$&&	1-2	$\sigma$&&	0 \FiveStarOpen&&	1 \FiveStarOpen\\[.02cm]
 	\cirn{4}~\cirn{5}~\cirn{8}&&	-1.949&&0.381&&	0.814 	&&	9.1169 	&&	1.05$\TIMES10^{-2}$&&	2-3	$\sigma$&&	1 \FiveStarOpen&&	1 \FiveStarOpen\\[.02cm]
 	\cirn{4}~\cirn{5}~\cirn{9}&&	-0.942&&0.401&&	0.826 	&&	8.4675 	&&	1.45$\TIMES10^{-2}$&&	2-3	$\sigma$&&	3 \FiveStarOpen&&	3 \FiveStarOpen\\[.02cm]
 	\cirn{4}~\cirn{5}~\cirn{12}&&	-1.704&&0.131&&	0.191 	&&	4.2477 	&&	1.20$\TIMES10^{-1}$&&	1-2	$\sigma$&&	0 \FiveStarOpen&&	0 \FiveStarOpen\\[.02cm]
 	\cirn{4}~\cirn{5}~\cirn{15}&&	-1.704&&0.131&&	0.191 	&&	4.2477 	&&	1.20$\TIMES10^{-1}$&&	1-2	$\sigma$&&	0 \FiveStarOpen&&	0 \FiveStarOpen\\[.02cm]
 	\cirn{4}~\cirn{7}~\cirn{8}&&	-1.704&&0.153&&	0.253 	&&	5.4225 	&&	6.65$\TIMES10^{-2}$&&	1-2	$\sigma$&&	0 \FiveStarOpen&&	1 \FiveStarOpen\\[.02cm]
 	\cirn{4}~\cirn{7}~\cirn{9}&&	-1.537&&0.163&&	0.303 	&&	6.9161 	&&	3.15$\TIMES10^{-2}$&&	2-3	$\sigma$&&	0 \FiveStarOpen&&	1 \FiveStarOpen\\[.02cm]
 	\cirn{4}~\cirn{7}~\cirn{12}&&	-1.686&&0.102&&	0.102 	&&	0.1161 	&&	9.44$\TIMES10^{-1}$&&	$<$1$\sigma$&&	0 \FiveStarOpen&&	0 \FiveStarOpen\\[.02cm]
 	\cirn{4}~\cirn{7}~\cirn{15}&&	-1.686&&0.102&&	0.102 	&&	0.1161 	&&	9.44$\TIMES10^{-1}$&&	$<$1$\sigma$&&	0 \FiveStarOpen&&	0 \FiveStarOpen\\[.02cm]
 	\cirn{4}~\cirn{8}~\cirn{9}&&	-1.385&&0.358&&	0.892 	&&	12.3840 &&	2.05$\TIMES10^{-3}$&&	3-4	$\sigma$&&	1 \FiveStarOpen&&	2 \FiveStarOpen\\[.02cm]
 	\cirn{4}~\cirn{8}~\cirn{12}&&	-1.744&&0.135&&	0.218 	&&	5.2530 	&&	7.23$\TIMES10^{-2}$&&	1-2	$\sigma$&&	0 \FiveStarOpen&&	0 \FiveStarOpen\\[.02cm]
 	\cirn{4}~\cirn{8}~\cirn{15}&&	-1.744&&0.135&&	0.218 	&&	5.2530 	&&	7.23$\TIMES10^{-2}$&&	1-2	$\sigma$&&	0 \FiveStarOpen&&	0 \FiveStarOpen\\[.02cm]
 	\cirn{4}~\cirn{9}~\cirn{12}&&	-1.642&&0.121&&	0.237 	&&	7.6432 	&&	2.19$\TIMES10^{-2}$&&	2-3	$\sigma$&&	0 \FiveStarOpen&&	1 \FiveStarOpen\\[.02cm]
 	\cirn{4}~\cirn{9}~\cirn{15}&&	-1.642&&0.121&&	0.237 	&&	7.6432 	&&	2.19$\TIMES10^{-2}$&&	2-3	$\sigma$&&	0 \FiveStarOpen&&	1 \FiveStarOpen\\[.02cm]
 	\cirn{4}~\cirn{12}~\cirn{15}&&-1.709&&0.095&&	0.095 	&&	0.0669 	&&	9.67$\TIMES10^{-1}$&&	$<$1$\sigma$&&	0 \FiveStarOpen&&	0 \FiveStarOpen\\[.02cm]
 	\cirn{5}~\cirn{6}~\cirn{9}&&	1.172&&	0.122&&	0.202 	&&	5.4673 	&&	6.50$\TIMES10^{-2}$&&	1-2	$\sigma$&&	1 \FiveStarOpen&&	2 \FiveStarOpen\\[.02cm]
 	\cirn{5}~\cirn{6}~\cirn{13}&&	1.523&&	0.114&&	0.413 	&&	26.1617 &&	2.08$\TIMES10^{-6}$&&	4-5	$\sigma$&&	0 \FiveStarOpen&&	1 \FiveStarOpen\\[.02cm]
 	\cirn{5}~\cirn{6}~\cirn{14}&&	1.523&&	0.114&&	0.413 	&&	26.1617 &&	2.08$\TIMES10^{-6}$&&	4-5	$\sigma$&&	0 \FiveStarOpen&&	1 \FiveStarOpen\\[.02cm]
 	\cirn{5}~\cirn{7}~\cirn{8}&&	-1.718&&0.163&&	0.355 	&&	9.5417 	&&	8.47$\TIMES10^{-3}$&&	2-3	$\sigma$&&	0 \FiveStarOpen&&	1 \FiveStarOpen\\[.02cm]
 	\cirn{5}~\cirn{7}~\cirn{9}&&	-1.527&&0.175&&	0.412 	&&	11.0753 &&	3.94$\TIMES10^{-3}$&&	2-3	$\sigma$&&	0 \FiveStarOpen&&	1 \FiveStarOpen\\[.02cm]
 	\cirn{5}~\cirn{7}~\cirn{12}&&	-1.691&&0.105&&	0.153 	&&	4.2574 	&&	1.19$\TIMES10^{-1}$&&	1-2	$\sigma$&&	0 \FiveStarOpen&&	0 \FiveStarOpen\\[.02cm]
 	\cirn{5}~\cirn{7}~\cirn{15}&&	-1.691&&0.105&&	0.153 	&&	4.2574 	&&	1.19$\TIMES10^{-1}$&&	1-2	$\sigma$&&	0 \FiveStarOpen&&	0 \FiveStarOpen\\[.02cm]
 	\cirn{5}~\cirn{8}~\cirn{9}&&	-1.111&&0.543&&	1.541 	&&	16.1149 &&	3.17$\TIMES10^{-4}$&&	3-4	$\sigma$&&	2 \FiveStarOpen&&	2 \FiveStarOpen\\[.02cm]
 	\cirn{5}~\cirn{8}~\cirn{12}&&	-1.761&&0.144&&	0.310 	&&	9.3076 	&&	9.53$\TIMES10^{-3}$&&	2-3	$\sigma$&&	0 \FiveStarOpen&&	1 \FiveStarOpen\\[.02cm]
 	\cirn{5}~\cirn{8}~\cirn{15}&&	-1.761&&0.144&&	0.310 	&&	9.3076 	&&	9.53$\TIMES10^{-3}$&&	2-3	$\sigma$&&	0 \FiveStarOpen&&	1 \FiveStarOpen\\[.02cm]
 	\cirn{5}~\cirn{9}~\cirn{12}&&	-1.645&&0.126&&	0.307 	&&	11.8146 &&	2.72$\TIMES10^{-3}$&&	2-3	$\sigma$&&	0 \FiveStarOpen&&	1 \FiveStarOpen\\[.02cm]
 	\cirn{5}~\cirn{9}~\cirn{13}&&	2.287&&	0.197&&	0.435 	&&	9.7314 	&&	7.71$\TIMES10^{-3}$&&	2-3	$\sigma$&&	0 \FiveStarOpen&&	1 \FiveStarOpen\\[.02cm]
 	\cirn{5}~\cirn{9}~\cirn{14}&&	2.287&&	0.197&&	0.435 	&&	9.7314 	&&	7.71$\TIMES10^{-3}$&&	2-3	$\sigma$&&	0 \FiveStarOpen&&	1 \FiveStarOpen\\[.02cm]
 	\cirn{5}~\cirn{9}~\cirn{15}&&	-1.645&&0.126&&	0.307 	&&	11.8146 &&	2.72$\TIMES10^{-3}$&&	2-3	$\sigma$&&	0 \FiveStarOpen&&	1 \FiveStarOpen\\[.02cm]
 	\cirn{5}~\cirn{12}~\cirn{15}&&-1.714&&0.098&&	0.142 	&&	4.1839 	&&	1.23$\TIMES10^{-1}$&&	1-2	$\sigma$&&	0 \FiveStarOpen&&	0 \FiveStarOpen\\[.02cm]
 	\cirn{5}~\cirn{13}~\cirn{14}&&2.386&&	0.134&&	0.193 	&&	4.1839 	&&	1.23$\TIMES10^{-1}$&&	1-2	$\sigma$&&	0 \FiveStarOpen&&	0 \FiveStarOpen\\[.02cm]
 	\cirn{6}~\cirn{9}~\cirn{13}&&	1.503&&	0.113&&	0.395 	&&	24.1799 &&	5.62$\TIMES10^{-6}$&&	4-5	$\sigma$&&	0 \FiveStarOpen&&	1 \FiveStarOpen\\[.02cm]
 	\cirn{6}~\cirn{9}~\cirn{14}&&	1.503&&	0.113&&	0.395 	&&	24.1799 &&	5.62$\TIMES10^{-6}$&&	4-5	$\sigma$&&	0 \FiveStarOpen&&	1 \FiveStarOpen\\[.02cm]
 	\cirn{6}~\cirn{13}~\cirn{14}&&1.711&&	0.101&&	0.419 	&&	34.3870 &&	3.41$\TIMES10^{-8}$&&	$>$5$\sigma$&&	0 \FiveStarOpen&&	1 \FiveStarOpen\\[.02cm]
 	\cirn{7}~\cirn{8}~\cirn{9}&&	-1.604&&0.166&&	0.422 	&&	12.8434 &&	1.63$\TIMES10^{-3}$&&	3-4	$\sigma$&&	0 \FiveStarOpen&&	1 \FiveStarOpen\\[.02cm]
 	\cirn{7}~\cirn{8}~\cirn{12}&&	-1.716&&0.106&&	0.173 	&&	5.3578 	&&	6.86$\TIMES10^{-2}$&&	1-2	$\sigma$&&	0 \FiveStarOpen&&	0 \FiveStarOpen\\[.02cm]
 	\cirn{7}~\cirn{8}~\cirn{15}&&	-1.716&&0.106&&	0.173 	&&	5.3578 	&&	6.86$\TIMES10^{-2}$&&	1-2	$\sigma$&&	0 \FiveStarOpen&&	0 \FiveStarOpen\\[.02cm]
 	\cirn{7}~\cirn{9}~\cirn{12}&&	-1.648&&0.102&&	0.199 	&&	7.6314 	&&	2.20$\TIMES10^{-2}$&&	2-3	$\sigma$&&	0 \FiveStarOpen&&	0 \FiveStarOpen\\[.02cm]
 	\cirn{7}~\cirn{9}~\cirn{15}&&	-1.648&&0.102&&	0.199 	&&	7.6314 	&&	2.20$\TIMES10^{-2}$&&	2-3	$\sigma$&&	0 \FiveStarOpen&&	0 \FiveStarOpen\\[.02cm]
 	\cirn{7}~\cirn{12}~\cirn{15}&&-1.699&&0.083&&	0.083 	&&	0.0919 	&&	9.55$\TIMES10^{-1}$&&	$<$1$\sigma$&&	0 \FiveStarOpen&&	0 \FiveStarOpen\\[.02cm]
 	\cirn{8}~\cirn{12}~\cirn{15}&&-1.737&&0.100&&	0.161 	&&	5.1804 	&&	7.50$\TIMES10^{-2}$&&	1-2	$\sigma$&&	0 \FiveStarOpen&&	0 \FiveStarOpen\\[.02cm]
 	\cirn{9}~\cirn{12}~\cirn{15}&&-1.676&&0.093&&	0.184 	&&	7.7679 	&&	2.06$\TIMES10^{-2}$&&	2-3	$\sigma$&&	0 \FiveStarOpen&&	0 \FiveStarOpen\\[.02cm]
 	\cirn{9}~\cirn{13}~\cirn{14}&&2.339&&	0.137&&	0.230 	&&	5.6794 	&&	5.84$\TIMES10^{-2}$&&	1-2	$\sigma$&&	0 \FiveStarOpen&&	0 \FiveStarOpen\\[.02cm]
\end{longtable*}

Self-consistent cases are not necessarily the majority, and the numbers of them are only 77 in NH and 61 in IH. As is the same with bipair combination, the number in NH exceeds that in IH slightly. On the other hand, cases over 3\FiveStarOpen~level have a number of 130 in NH and 146 in IH, nearly half of the total. There are 39 (NH) and 16 (IH) cases satisfying the two conditions simultaneously.

Similarly, most of the cases satisfying the two conditions are compatible with the maximal {\it CP} violation in $3\sigma$ range. Exceptions are \cirn{\small{2}}~\cirn{\small{4}}~\cirn{\small{5}} and \cirn{\small{2}}~\cirn{\small{4}}~\cirn{\small{9}} in IH and with unscaled error. Especially, with unscaled errors, 21 of all 39 cases in NH are compatible with the maximal {\it CP} violation within $1\sigma$ range. When errors are scaled, the number increases to 25. The detailed results are listed in Table~\ref{tab:CPV_NH} (NH) and Table~\ref{tab:CPV_IH} (IH) in Sec.~\ref{sec:mCPV}.

\section{checking the maximal {\it CP} violation} \label{sec:mCPV}

In this section we compare constraints by the pairs to the maximal {\it CP} violation. We include all cases that are self-consistent and over 3\FiveStarOpen~level and discuss separately in NH and IH.

For each case, we list the deviation from the maximal {\it CP} violation for both unscaled and scaled errors. The results are listed in Table~\ref{tab:CPV_NH} (NH) and Table~\ref{tab:CPV_IH} (IH).

\begin{longtable*}{ccccccccccc}
\caption{\label{tab:CPV_NH}Deviations from the maximal {\it CP} violation in NH.}\\
\hline\\[-3mm]
\hline\\[-.3cm]
\multirow{2}{*}{Pairs included}&&\multirow{2}{*}{Exclusion level of self-consistency}&&\multicolumn{3}{c}{Natural limit consistency}&&\multicolumn{3}{c}{Deviation from the maximal {\it CP} violation}\\
\cline{5-7}\cline{9-11}\\[-.27cm]
&&&&Unscaled&&Scaled&&Unscaled&&Scaled\\
\\[-.4cm] \hline
\endhead
\hline \multicolumn{11  }{r}{{Continued on next page}} \\ \hline \hline
\endfoot
\hline \hline
\endlastfoot
 	 	\cirn{2}~\cirn{3}~\cirn{4}	&&	1-2	$\sigma$&&	4	\FiveStarOpen&&	4	\FiveStarOpen&&	     $<$1	$\sigma$&&	     $<$1	$\sigma$\\
 	 	\cirn{1}~\cirn{2}~\cirn{5}	&&	$<$1$\sigma$&&	5	\FiveStarOpen&&	5	\FiveStarOpen&&	     $<$1	$\sigma$&&	     $<$1	$\sigma$\\
 	 	\cirn{1}~\cirn{2}~\cirn{3}~\cirn{4}~\cirn{5}&&	1-2	$\sigma$&&	5\FiveStarOpen&&	5\FiveStarOpen&&$<$1$\sigma$&&	     $<$1	$\sigma$\\
        \cirn{1}            &&  $\cdot\cdot\cdot$ &&  5	\FiveStarOpen&&   $\cdot\cdot\cdot$ &&	     $<$1$\sigma$&& $\cdot\cdot\cdot$\\
        \cirn{2}            &&  $\cdot\cdot\cdot$ &&  4	\FiveStarOpen&&   $\cdot\cdot\cdot$ &&	     $<$1$\sigma$&& $\cdot\cdot\cdot$\\
        \cirn{4}            &&  $\cdot\cdot\cdot$ &&  4	\FiveStarOpen&&   $\cdot\cdot\cdot$ &&	     $<$1$\sigma$&& $\cdot\cdot\cdot$\\
        \cirn{8}            &&  $\cdot\cdot\cdot$ &&  3	\FiveStarOpen&&   $\cdot\cdot\cdot$ &&      1-2 $\sigma$&& $\cdot\cdot\cdot$\\
 	 	\cirn{1}~\cirn{2}	&&	1-2	$\sigma$&&	5	\FiveStarOpen&&	5	\FiveStarOpen&&	     $<$1$\sigma$&&	     $<$1	$\sigma$\\
 	 	\cirn{1}~\cirn{3}	&&	2-3	$\sigma$&&	5	\FiveStarOpen&&	4	\FiveStarOpen&&	     $<$1$\sigma$&&	     $<$1	$\sigma$\\
 	 	\cirn{1}~\cirn{4}	&&	$<$1$\sigma$&&	5	\FiveStarOpen&&	5	\FiveStarOpen&&	     1-2 $\sigma$&&	     1-2	$\sigma$\\
 	 	\cirn{1}~\cirn{5}	&&	$<$1$\sigma$&&	5	\FiveStarOpen&&	5	\FiveStarOpen&&	     1-2 $\sigma$&&	     1-2	$\sigma$\\
 	 	\cirn{1}~\cirn{8}	&&	$<$1$\sigma$&&	5	\FiveStarOpen&&	5	\FiveStarOpen&&	     1-2 $\sigma$&&	     1-2	$\sigma$\\
 	 	\cirn{1}~\cirn{9}	&&	2-3	$\sigma$&&	5	\FiveStarOpen&&	4	\FiveStarOpen&&	     $<$1$\sigma$&&	     $<$1	$\sigma$\\
 	 	\cirn{2}~\cirn{3}	&&	1-2	$\sigma$&&	4	\FiveStarOpen&&	4	\FiveStarOpen&&	     2-3 $\sigma$&&	     2-3	$\sigma$\\
 	 	\cirn{2}~\cirn{4}	&&	1-2	$\sigma$&&	5	\FiveStarOpen&&	4	\FiveStarOpen&&	     $<$1$\sigma$&&	     $<$1	$\sigma$\\
 	 	\cirn{2}~\cirn{5}	&&	$<$1$\sigma$&&	4	\FiveStarOpen&&	4	\FiveStarOpen&&	     1-2 $\sigma$&&	     1-2	$\sigma$\\
 	 	\cirn{2}~\cirn{8}	&&	1-2	$\sigma$&&	4	\FiveStarOpen&&	4	\FiveStarOpen&&	     $<$1$\sigma$&&	     $<$1	$\sigma$\\
 	 	\cirn{2}~\cirn{9}	&&	2-3	$\sigma$&&	4	\FiveStarOpen&&	3	\FiveStarOpen&&	     2-3 $\sigma$&&	     1-2	$\sigma$\\
 	 	\cirn{3}~\cirn{4}	&&	2-3	$\sigma$&&	4	\FiveStarOpen&&	4	\FiveStarOpen&&	     $<$1$\sigma$&&	     $<$1	$\sigma$\\
 	 	\cirn{3}~\cirn{8}	&&	2-3	$\sigma$&&	4	\FiveStarOpen&&	3	\FiveStarOpen&&	     $<$1$\sigma$&&	     $<$1	$\sigma$\\
 	 	\cirn{4}~\cirn{5}	&&	$<$1$\sigma$&&	4	\FiveStarOpen&&	4	\FiveStarOpen&&	     $<$1$\sigma$&&	     $<$1	$\sigma$\\
 	 	\cirn{4}~\cirn{8}	&&	$<$1$\sigma$&&	4	\FiveStarOpen&&	4	\FiveStarOpen&&	     1-2 $\sigma$&&	     1-2	$\sigma$\\
 	 	\cirn{4}~\cirn{9}	&&	2-3	$\sigma$&&	4	\FiveStarOpen&&	4	\FiveStarOpen&&	     $<$1$\sigma$&&	     $<$1	$\sigma$\\
 	 	\cirn{5}~\cirn{8}	&&	$<$1$\sigma$&&	3	\FiveStarOpen&&	3	\FiveStarOpen&&	     $<$1$\sigma$&&	     $<$1	$\sigma$\\
 	 	\cirn{8}~\cirn{9}	&&	2-3	$\sigma$&&	4	\FiveStarOpen&&	3	\FiveStarOpen&&	     $<$1$\sigma$&&	     $<$1	$\sigma$\\
 	 	\cirn{1}~\cirn{2}~\cirn{3}	&&	1-2	$\sigma$&&	5	\FiveStarOpen&&	4	\FiveStarOpen&&	     $<$1$\sigma$&&	     $<$1	$\sigma$\\
 	 	\cirn{1}~\cirn{2}~\cirn{4}	&&	$<$1$\sigma$&&	5	\FiveStarOpen&&	5	\FiveStarOpen&&	     1-2 $\sigma$&&	     1-2	$\sigma$\\
 	 	\cirn{1}~\cirn{2}~\cirn{8}	&&	1-2	$\sigma$&&	5	\FiveStarOpen&&	5	\FiveStarOpen&&	     1-2 $\sigma$&&	     1-2	$\sigma$\\
 	 	\cirn{1}~\cirn{2}~\cirn{9}	&&	2-3	$\sigma$&&	5	\FiveStarOpen&&	4	\FiveStarOpen&&	     $<$1$\sigma$&&	     $<$1	$\sigma$\\
 	 	\cirn{1}~\cirn{3}~\cirn{4}	&&	1-2	$\sigma$&&	5	\FiveStarOpen&&	5	\FiveStarOpen&&	     $<$1$\sigma$&&	     $<$1	$\sigma$\\
 	 	\cirn{1}~\cirn{3}~\cirn{5}	&&	1-2	$\sigma$&&	5	\FiveStarOpen&&	4	\FiveStarOpen&&	     $<$1$\sigma$&&	     $<$1	$\sigma$\\
 	 	\cirn{1}~\cirn{3}~\cirn{8}	&&	2-3	$\sigma$&&	5	\FiveStarOpen&&	4	\FiveStarOpen&&	     $<$1$\sigma$&&	     $<$1	$\sigma$\\
 	 	\cirn{1}~\cirn{3}~\cirn{9}	&&	2-3	$\sigma$&&	5	\FiveStarOpen&&	4	\FiveStarOpen&&	     $<$1$\sigma$&&	     $<$1	$\sigma$\\
 	 	\cirn{1}~\cirn{4}~\cirn{5}	&&	$<$1$\sigma$&&	5	\FiveStarOpen&&	5	\FiveStarOpen&&	     1-2 $\sigma$&&	     1-2	$\sigma$\\
 	 	\cirn{1}~\cirn{4}~\cirn{8}	&&	$<$1$\sigma$&&	5	\FiveStarOpen&&	5	\FiveStarOpen&&	     1-2 $\sigma$&&	     1-2	$\sigma$\\
 	 	\cirn{1}~\cirn{4}~\cirn{9}	&&	2-3	$\sigma$&&	5	\FiveStarOpen&&	4	\FiveStarOpen&&	     1-2 $\sigma$&&	     $<$1	$\sigma$\\
 	 	\cirn{1}~\cirn{5}~\cirn{8}	&&	$<$1$\sigma$&&	5	\FiveStarOpen&&	5	\FiveStarOpen&&	     1-2 $\sigma$&&	     1-2	$\sigma$\\
 	 	\cirn{1}~\cirn{5}~\cirn{9}	&&	2-3	$\sigma$&&	5	\FiveStarOpen&&	4	\FiveStarOpen&&	     $<$1$\sigma$&&	     $<$1	$\sigma$\\
 	 	\cirn{1}~\cirn{8}~\cirn{9}	&&	2-3	$\sigma$&&	5	\FiveStarOpen&&	4	\FiveStarOpen&&	     1-2 $\sigma$&&	     $<$1	$\sigma$\\
	 	\cirn{2}~\cirn{3}~\cirn{5}	&&	$<$1$\sigma$&&	4	\FiveStarOpen&&	4	\FiveStarOpen&&	     2-3 $\sigma$&&	     2-3	$\sigma$\\
 	 	\cirn{2}~\cirn{3}~\cirn{8}	&&	1-2	$\sigma$&&	4	\FiveStarOpen&&	4	\FiveStarOpen&&	     1-2 $\sigma$&&	     $<$1	$\sigma$\\
 	 	\cirn{2}~\cirn{3}~\cirn{9}	&&	1-2	$\sigma$&&	4	\FiveStarOpen&&	3	\FiveStarOpen&&	     2-3 $\sigma$&&	     1-2	$\sigma$\\
 	 	\cirn{2}~\cirn{4}~\cirn{5}	&&	$<$1$\sigma$&&	5	\FiveStarOpen&&	4	\FiveStarOpen&&	     $<$1$\sigma$&&	     $<$1	$\sigma$\\
 	 	\cirn{2}~\cirn{4}~\cirn{8}	&&	1-2	$\sigma$&&	4	\FiveStarOpen&&	4	\FiveStarOpen&&	     $<$1$\sigma$&&	     $<$1	$\sigma$\\
 	 	\cirn{2}~\cirn{4}~\cirn{9}	&&	2-3	$\sigma$&&	4	\FiveStarOpen&&	4	\FiveStarOpen&&	     $<$1$\sigma$&&	     $<$1	$\sigma$\\
 	 	\cirn{2}~\cirn{5}~\cirn{8}	&&	1-2	$\sigma$&&	4	\FiveStarOpen&&	4	\FiveStarOpen&&	     $<$1$\sigma$&&	     $<$1	$\sigma$\\
 	 	\cirn{2}~\cirn{5}~\cirn{9}	&&	1-2	$\sigma$&&	4	\FiveStarOpen&&	3	\FiveStarOpen&&	     2-3 $\sigma$&&	     1-2	$\sigma$\\
 	 	\cirn{2}~\cirn{8}~\cirn{9}	&&	2-3	$\sigma$&&	4	\FiveStarOpen&&	3	\FiveStarOpen&&	     $<$1$\sigma$&&	     $<$1	$\sigma$\\
 	 	\cirn{3}~\cirn{4}~\cirn{5}	&&	1-2	$\sigma$&&	4	\FiveStarOpen&&	4	\FiveStarOpen&&	     $<$1$\sigma$&&	     $<$1	$\sigma$\\
 	 	\cirn{3}~\cirn{4}~\cirn{8}	&&	2-3	$\sigma$&&	5	\FiveStarOpen&&	4	\FiveStarOpen&&	     $<$1$\sigma$&&	     $<$1	$\sigma$\\
 	 	\cirn{3}~\cirn{4}~\cirn{9}	&&	2-3	$\sigma$&&	4	\FiveStarOpen&&	3	\FiveStarOpen&&	     $<$1$\sigma$&&	     $<$1	$\sigma$\\
 	 	\cirn{3}~\cirn{5}~\cirn{8}	&&	2-3	$\sigma$&&	4	\FiveStarOpen&&	3	\FiveStarOpen&&	     $<$1$\sigma$&&	     $<$1	$\sigma$\\
 	 	\cirn{3}~\cirn{8}~\cirn{9}	&&	2-3	$\sigma$&&	3	\FiveStarOpen&&	3	\FiveStarOpen&&	     1-2 $\sigma$&&	     $<$1	$\sigma$\\
 	 	\cirn{4}~\cirn{5}~\cirn{8}	&&	$<$1$\sigma$&&	4	\FiveStarOpen&&	4	\FiveStarOpen&&	     1-2 $\sigma$&&	     1-2	$\sigma$\\
 	 	\cirn{4}~\cirn{5}~\cirn{9}	&&	2-3	$\sigma$&&	4	\FiveStarOpen&&	4	\FiveStarOpen&&	     $<$1$\sigma$&&	     $<$1	$\sigma$\\
 	 	\cirn{4}~\cirn{8}~\cirn{9}	&&	2-3	$\sigma$&&	4	\FiveStarOpen&&	4	\FiveStarOpen&&	     $<$1$\sigma$&&	     $<$1	$\sigma$\\
 	 	\cirn{5}~\cirn{8}~\cirn{9}	&&	2-3	$\sigma$&&	4	\FiveStarOpen&&	3	\FiveStarOpen&&	     $<$1$\sigma$&&	     $<$1	$\sigma$\\
\end{longtable*}

\begin{longtable*}{ccccccccccc}
\caption{\label{tab:CPV_IH}Deviations from the maximal {\it CP} violation in IH.}\\
\hline\\[-3mm]
\hline\\[-.3cm]
\multirow{2}{*}{Pairs included}&&\multirow{2}{*}{Exclusion level of self-consistency}&&\multicolumn{3}{c}{Natural limit consistency}&&\multicolumn{3}{c}{Deviation from the maximal {\it CP} violation}\\
\cline{5-7}\cline{9-11}\\[-.27cm]
&&&&Unscaled&&Scaled&&Unscaled&&Scaled\\
\\[-.4cm] \hline
\endhead
\hline \multicolumn{11  }{r}{{Continued on next page}} \\ \hline \hline
\endfoot
\hline \hline
\endlastfoot
 	 	\cirn{2}~\cirn{3}~\cirn{4}	&&	2-3	$\sigma$&&	4	\FiveStarOpen&&	3	\FiveStarOpen&&	     2-3	$\sigma$&&	     1-2	$\sigma$\\
 	 	\cirn{1}~\cirn{2}~\cirn{5}	&&	2-3	$\sigma$&&	5	\FiveStarOpen&&	4	\FiveStarOpen&&	     1-2	$\sigma$&&	     $<$1	$\sigma$\\
        \cirn{1}            &&  $\cdot\cdot\cdot$ &&  5	\FiveStarOpen&&   $\cdot\cdot\cdot$  &&      2-3	$\sigma$&& $\cdot\cdot\cdot$\\
        \cirn{2}            &&  $\cdot\cdot\cdot$ &&  3	\FiveStarOpen&&   $\cdot\cdot\cdot$  &&	     2-3	$\sigma$&& $\cdot\cdot\cdot$\\
        \cirn{3}            &&  $\cdot\cdot\cdot$ &&  4	\FiveStarOpen&&   $\cdot\cdot\cdot$  &&	     $<$1	$\sigma$&& $\cdot\cdot\cdot$\\
        \cirn{9}            &&  $\cdot\cdot\cdot$ &&  3	\FiveStarOpen&&   $\cdot\cdot\cdot$  &&      $<$1	$\sigma$&& $\cdot\cdot\cdot$\\
 	 	\cirn{1}~\cirn{2}	&&	     2-3	$\sigma$&&	5	\FiveStarOpen&&	4	\FiveStarOpen&&	     1-2	$\sigma$&&	     $<$1	$\sigma$\\
 	 	\cirn{1}~\cirn{3}	&&	     $<$1	$\sigma$&&	5	\FiveStarOpen&&	5	\FiveStarOpen&&	     2-3	$\sigma$&&	     2-3	$\sigma$\\
 	 	\cirn{1}~\cirn{5}	&&	     2-3	$\sigma$&&	5	\FiveStarOpen&&	4	\FiveStarOpen&&	     2-3	$\sigma$&&	     1-2	$\sigma$\\
 	 	\cirn{1}~\cirn{9}	&&	     $<$1	$\sigma$&&	5	\FiveStarOpen&&	5	\FiveStarOpen&&	     2-3	$\sigma$&&	     2-3	$\sigma$\\
 	 	\cirn{2}~\cirn{3}	&&	     1-2	$\sigma$&&	4	\FiveStarOpen&&	4	\FiveStarOpen&&	     1-2	$\sigma$&&	     $<$1	$\sigma$\\
 	 	\cirn{2}~\cirn{4}	&&	     1-2	$\sigma$&&	3	\FiveStarOpen&&	3	\FiveStarOpen&&	     3-4	$\sigma$&&	     2-3	$\sigma$\\
 	 	\cirn{2}~\cirn{5}	&&	     2-3	$\sigma$&&	3	\FiveStarOpen&&	3	\FiveStarOpen&&	     2-3	$\sigma$&&	     1-2	$\sigma$\\
 	 	\cirn{2}~\cirn{9}	&&	     1-2	$\sigma$&&	4	\FiveStarOpen&&	4	\FiveStarOpen&&	     1-2	$\sigma$&&	     1-2	$\sigma$\\
 	 	\cirn{3}~\cirn{4}	&&	     2-3	$\sigma$&&	4	\FiveStarOpen&&	3	\FiveStarOpen&&	     1-2	$\sigma$&&	     $<$1	$\sigma$\\
 	 	\cirn{3}~\cirn{5}	&&	     2-3	$\sigma$&&	4	\FiveStarOpen&&	4	\FiveStarOpen&&	     $<$1	$\sigma$&&	     $<$1	$\sigma$\\
 	 	\cirn{3}~\cirn{9}	&&	     $<$1	$\sigma$&&	4	\FiveStarOpen&&	4	\FiveStarOpen&&	     $<$1	$\sigma$&&	     $<$1	$\sigma$\\
 	 	\cirn{4}~\cirn{9}	&&	     2-3	$\sigma$&&	3	\FiveStarOpen&&	3	\FiveStarOpen&&	     2-3	$\sigma$&&	     1-2	$\sigma$\\
 	 	\cirn{5}~\cirn{9}	&&	     2-3	$\sigma$&&	4	\FiveStarOpen&&	3	\FiveStarOpen&&	     $<$1	$\sigma$&&	     $<$1	$\sigma$\\
 	 	\cirn{1}~\cirn{2}~\cirn{3}	&&	2-3	$\sigma$&&	5	\FiveStarOpen&&	5	\FiveStarOpen&&	     1-2	$\sigma$&&	     $<$1	$\sigma$\\
 	 	\cirn{1}~\cirn{2}~\cirn{9}	&&	2-3	$\sigma$&&	5	\FiveStarOpen&&	4	\FiveStarOpen&&	     1-2	$\sigma$&&	     $<$1	$\sigma$\\
 	 	\cirn{1}~\cirn{3}~\cirn{5}	&&	1-2	$\sigma$&&	5	\FiveStarOpen&&	5	\FiveStarOpen&&	     2-3	$\sigma$&&	     1-2	$\sigma$\\
 	 	\cirn{1}~\cirn{3}~\cirn{9}	&&	$<$1$\sigma$&&	5	\FiveStarOpen&&	5	\FiveStarOpen&&	     2-3	$\sigma$&&	     2-3	$\sigma$\\
 	 	\cirn{1}~\cirn{5}~\cirn{9}	&&	1-2	$\sigma$&&	5	\FiveStarOpen&&	5	\FiveStarOpen&&	     2-3	$\sigma$&&	     1-2	$\sigma$\\
 	 	\cirn{2}~\cirn{3}~\cirn{5}	&&	2-3	$\sigma$&&	4	\FiveStarOpen&&	4	\FiveStarOpen&&	     1-2	$\sigma$&&	     $<$1	$\sigma$\\
 	 	\cirn{2}~\cirn{3}~\cirn{9}	&&	1-2	$\sigma$&&	4	\FiveStarOpen&&	4	\FiveStarOpen&&	     1-2	$\sigma$&&	     1-2	$\sigma$\\
 	 	\cirn{2}~\cirn{4}~\cirn{5}	&&	1-2	$\sigma$&&	3	\FiveStarOpen&&	3	\FiveStarOpen&&	     3-4	$\sigma$&&	     2-3	$\sigma$\\
 	 	\cirn{2}~\cirn{4}~\cirn{9}	&&	1-2	$\sigma$&&	3	\FiveStarOpen&&	3	\FiveStarOpen&&	     3-4	$\sigma$&&	     2-3	$\sigma$\\
 	 	\cirn{2}~\cirn{5}~\cirn{9}	&&	1-2	$\sigma$&&	4	\FiveStarOpen&&	3	\FiveStarOpen&&	     1-2	$\sigma$&&	     1-2	$\sigma$\\
 	 	\cirn{3}~\cirn{4}~\cirn{5}	&&	2-3	$\sigma$&&	4	\FiveStarOpen&&	3	\FiveStarOpen&&	     1-2	$\sigma$&&	     $<$1	$\sigma$\\
 	 	\cirn{3}~\cirn{4}~\cirn{9}	&&	2-3	$\sigma$&&	4	\FiveStarOpen&&	4	\FiveStarOpen&&	     1-2	$\sigma$&&	     $<$1	$\sigma$\\
 	 	\cirn{3}~\cirn{5}~\cirn{9}	&&	1-2	$\sigma$&&	4	\FiveStarOpen&&	4	\FiveStarOpen&&	     $<$1	$\sigma$&&	     $<$1	$\sigma$\\
 	 	\cirn{4}~\cirn{5}~\cirn{9}	&&	2-3	$\sigma$&&	3	\FiveStarOpen&&	3	\FiveStarOpen&&	     2-3	$\sigma$&&	     1-2	$\sigma$\\
\end{longtable*}

As is mentioned in the previous section, combinations in NH tend to be slightly more self-consistent. All of the cases in NH are compatible with the maximal {\it CP} violation in $3\sigma$ range, while 3 exceptions are found in IH with errors unscaled. What is more, cases in NH seem to be more consistent with the maximal {\it CP} violation, since the majority of the cases (36 in 57 cases) deviate from the maximal {\it CP} violation within $1\sigma$ range, even when errors are not scaled. However, in IH only 6 in all 33 cases are compatible with the maximal {\it CP} violation in $1\sigma$ with unscaled errors. When errors are scaled, the number increases to 15.

\section{conclusions}\label{sec:conclusion}

From the examination of the pairs, we find that some of the pairs are not so consistent with the experimental results, and that some of the pairs are not consistent with each other when combined together. When seeking for new mixing patterns, these cases are not good choices. On the other hand, some of the cases agree with current experimental results and are self-consistent as well, such as the trimaximal mixing case and the $\mu$-$\tau$ symmetry. They can act as a starting point when constructing a new mixing pattern. While the first case of the bipair mixing (i.e., \cirn{\small{4}}~\cirn{\small{8}}) is a good choice only in NH and the second case \cirn{\small{3}}~\cirn{\small{9}} is good only in IH.

In addition, the examination provides information about the constraint on the {\it CP} phase by pairs. Especially, among cases that are both self-consistent and consistent with the natural limit, a majority of them are compatible with the maximal {\it CP} violation. It is necessary to point out that when deriving the range of the {\it CP} phase in Fig.~\ref{fig:delta}, we adopt the assumption of $\delta\in[0^\circ,180^\circ]$. When extending to $[-180^\circ,180^\circ]$, the results of $\delta\in[\delta_1,\delta_2]$ should also be extended to be $\delta\in[-\delta_2,-\delta_1]$ and $[\delta_1,\delta_2]$. Therefore, the results agree with the hint of the maximal CP violation with $\delta \sim -90^\circ$ from analysis in Ref.~\cite{analysis} and global fit results.

\begin{acknowledgments}
This work is supported by the Principal Fund for Undergraduate Research at Peking
University.
It is also supported by the National Natural Science Foundation of China (Grants No.~11035003 and No.~11120101004) and by the National Fund for Fostering
Talents of Basic Science (Grants No.~J1103205 and No. J1103206).
\end{acknowledgments}

\end{document}